\documentclass[twocolumn]{aastex631}

\hypersetup{linkcolor=red,citecolor=blue,filecolor=cyan,urlcolor=magenta}

\accepted{November 26th, 2025}
\submitjournal{ApJ}

\shorttitle{Star formation history of low-z AGN}

\usepackage{txfonts}
\usepackage{rotating}
\usepackage{subfigure}
\usepackage[T1]{fontenc}

\begin{document}

\title{Galaxy and Mass Assembly (GAMA): The Properties of Quasar Host Galaxies: Star Formation Histories and Stellar Populations.}

\correspondingauthor{Maria Stone}
\email{mbstone12@gmail.com}

\author[0000-0002-2931-0593]{Maria B. Stone}
\affiliation{Department of Physics and Astronomy,
Vesilinnantie 5, FI-20014, University of Turku, Finland}
\affiliation{Finnish Centre for Astronomy with ESO (FINCA), 
Vesilinnantie 5, FI-20014, University of Turku, Finland}
\affiliation{Institute for Space Astrophysics and Planetology (IAPS),
National Institute for Astrophysics (INAF),
via del Fosso del Cavaliere 100, 00133, Rome, Italy}

\author[0000-0003-1455-7339]{Roberto De Propris}
\affiliation{Finnish Centre for Astronomy with ESO (FINCA), 
Vesilinnantie 5, FI-20014, University of Turku, Finland}
\affiliation{Department of Physics and Astronomy, 
Botswana International University of Science and Technology (BIUST), 
Private Bag 16, Palapye, Botswana}

\author[0000-0002-7135-2842]{Clare Wethers}
\affiliation{Department of Space, Earth and Environment, 
Chalmers University of Technology, 
Onsala Space Observatory, 439 92, Onsala, Sweden}

\author[0000-0003-0133-7644]{Jari Kotilainen}
\affiliation{Finnish Centre for Astronomy with ESO (FINCA), 
Vesilinnantie 5, FI-20014, University of Turku, Finland}
\affiliation{Department of Physics and Astronomy,
Vesilinnantie 5, FI-20014, University of Turku, Finland}

\author[0000-0001-9738-3594]{Nischal Acharya}
\affiliation{Donostia International Physics Center (DIPC), 
Paseo Manuel de Lordizabal 4, E-20018 Donostia-San Sebastian, Spain}
%\nocollaboration{5}

\author[0000-0002-4884-6756]{Benne Holwerda}
\affiliation{Department of Physics and Astronomy,
University of Louisville, 
102 Natural Sciences Building, Louisville, KY, 40292, USA}

\author[0000-0002-6097-2747]{Andrew M. Hopkins}
\affiliation{School of Mathematical and Physical Sciences, 
12 Wally's Walk, Macquarie University, NSW 2109, Australia
}

\author[0000-0002-3963-3919]{Kevin Pimbblet}
\affiliation{E.A. Milne Centre, Faculty of Science and Engineering, University of Hull, Cottingham Road, Kingston-upon-Hull HU6 7RX, UK}; \affiliation{Centre of Excellence for Data Science, AI, and Modelling (DAIM), University of Hull, Cottingham Road, Kingston-upon-Hull, HU6 7RX, UK }

%\collaboration{20}{(GAMA Collaboration)}

\begin{abstract}

We investigated the star formation history and stellar populations of a sample of 205 Type~I quasar host galaxies (0.1$<$z$<$0.35) and compared with normal (non-active) galaxies of the same mass and redshift within the volume of the Galaxy and Mass Assembly (GAMA) redshift survey. 
We find that quasar host galaxies tend to be star-forming galaxies ($\sim$ 80\%) lying on the star-forming MS; the fraction of quasar host galaxies that are quiescent ($\sim$ 20\%) is lower than the fraction of quiescent galaxies in the comparison sample of normal galaxies (54\%). 
We find that the mean star formation rate of quasar host galaxies has increased over the past 100~Myr by a factor of 2--3, but these galaxies were star-forming at all times previously.  
Our data are more consistent with quasar activity originating together with an increase in the star formation rate of otherwise normal galaxies, similar to episodic star formation in normal spirals. 
We argue that this indicates that secular processes and minor mergers may be the favored triggers of nuclear activity in the local Universe.

\end{abstract}


\section{Introduction} \label{sec:intro}

All massive galaxies contain supermassive black holes (SMBH) in their nuclei (e.g.,
\citealt{Magorrian_1998,Silk_1998,Gebhardt_2000}). 
The mass of these black holes correlates tightly with the mass (or velocity dispersion) of their parent spheroids (e.g., \citealt{Gultekin_2009}), despite very large (several orders of magnitude) differences in the size and mass of these two components. 
This points to a mechanism of co-evolution between SMBHs and their parent galaxies (see \citealt{Kormendy_2013} for a review, however see \citealt{Jahnke_2011} for a critique) such as `quasar-mode' feedback (e.g., \citealt{Terrazas_2020}), where accretion of matter into a SMBH, leads to the triggering of an active galactic nucleus (AGN), and can inject enough energy into the galaxy to affect its star formation through heating, ionization by the ultraviolet flux, or mechanical input from jets (e.g., \citealt{Granato_2004,Zubovas_2013,Zubovas_2017,Fabian_2012,King_2015,Blandford_2019,
Trussler_2020}). However, these outflows and jets can also boost star formation (e.g., \citealt{Hopkins_2012,
Nayakshin_2012,Bieri_2016,Zubovas_2013,Zubovas_2017}) even form stars inside the outflow (e.g.,
\citealt{Ishibashi_2012,Zubovas_2013,El-Badry_2016,Wang_2018}). These mechanisms may also co-exist within the same object (e.g., \citealt{Cresci_2015,Shin_2019,Mandal_2021}). 

Previous studies (e.g., \citealt{Asari_2007,Sarzi_2007}) concluded that in 30--50\% of the cases, the AGN is associated with young stellar populations (i.e. older than a few 100~Myr). 
However, this could be due to the fact that both AGN and starbursts require gas to fuel them and not necessarily imply a causal relation. 
For example, the color and morphology of low redshift ($z < 0.3$) quasar host galaxies are not significantly different from those of normal (non-active) galaxies \citep{Bettoni_2015}, suggesting that nuclear activity does not affect the global properties of galaxies.
Literature results regarding the AGN-starburst connection are controversial. 
It is yet unclear whether AGN activity occurs together with star formation \citep{Kawakatu_2008}, follows it during a post-starburst phase \citep{Davies_2007,Riffel_2009,Riffel_2021}, or there is no association with recent star formation \citep{Sarzi_2007}. 
The observational situation is ambiguous, with several studies showing evidence for quenching \citep{Page_2012,Barger_2015,Shimizu_2015,Jin_2018,Stemo_2020}, enhanced star formation \citep{Lutz_2010,Mullaney_2012,Santini_2012,Jarvis_2020,Shangguan_2020,Xie_2021, Koutoulidis_2022}, or no significant effect \citep{Harrison_2012,Rosario_2012,Husemann_2014,Stanley_2015, Stanley_2017,Suh_2017,Woo_2017,Smirnova_2022}. 
One possibility is that these discrepancies reflect the different samples of AGN, choice of comparison galaxies, and even evolutionary effects within the host galaxy population as a function of redshift.
Similarly, earlier claims have been made that AGN hosts are dominated by old stellar populations \citep{McLure_1999,Nolan_2001} and that they contain younger stars \citep{Kauffmann_2003,Jin_2018}.
On the one hand, we expect that AGN should accelerate quenching \citep{Gofford_2015,Hopkins_2016}, but this may also take place over longer timescales \citep{Rembold_2017,Sanchez_2018}. 
It is possible that these conflicting results reflect the influence of selection effects, different environments, and disparate methods to derive star formation histories (SFHs) for AGN host galaxies and the comparison systems. 
In \cite{Graham_2024} mergers drive morphological transformation, not AGN feedback per se.

With this in mind, in this study we set out to compare the star formation properties of a sample of Type~I AGN hosts and of a comparison sample of normal (non-active or without quasar activity) galaxies (CSNG) selected within the same volume.
The samples are matched in the stellar mass distribution and redshift. 
Our purpose is to compare the SFHs of AGN host galaxies and those of normal galaxies to understand the role of AGN activity in shaping the stellar populations and SFHs of galaxies. 
We used the catalog of low-redshift Type I quasars by \cite{Wethers_2022}, which assembled all known Type~I quasars from the recent compilation of quasar catalogs in \cite{Gattano_2018} within the three equatorial fields of the Galaxy and Mass Assembly (GAMA) spectroscopic survey \citep{Liske_2015,Driver_2018,Driver_2022}. 
We also select 200 sets of normal galaxies with the same redshift and mass distribution as the quasars to act as a comparison sample. 
This should lessen many of the selection effects that affect previous studies (such as any possible evolutionary issues, for example, since all objects are observed at the same epoch within a few hundred Myr).
The star formation properties for CSNG are pulled from GAMA archive.
For quasar hosts, we use the \textsc{cigale} spectral energy distribution (SED) analysis tool to derive estimates of star formation properties, employing the most up-to-date AGN model available.

The structure of this paper is as follows: in \S~\ref{sec:data} we discuss the data samples and the property estimates pulled from various archival databases. 
In \S~\ref{sec:star_properties}, we detail how we estimate stellar population parameters for the quasar host galaxies. 
In \S~\ref{sec:results} we present our main results, comparing star formation rates (SFR) and SFHs for AGN hosts and normal galaxies. 
We discuss our results in \S~\ref{sec:discussion} in the light of models for AGN formation and feedback. 
Conclusions are presented in \S~\ref{sec:conclusions}. 
Throughout this Paper we use cosmological parameters from \cite{Planck_VI}.

\section{The samples}
\label{sec:data}

Based on the data provided by the GAMA spectroscopic survey, first the sample of quasars is assembled (\S~\ref{subsec:quasar_sample}) consisting of all known bright low-\textit{z} Type I AGN which are located in the GAMA equatorial footprint. Observational properties of the quasar sample are given in \S~\ref{subsec:properties}, based on the existing surveys.
Since there are many different types of AGN, these additional properties are given to indicate that overall the quasar sample in this work consists of typical low-redshift Type I AGN; our quasar sample is representative of that group of quasars. 

Next, we extract the comparison sample of normal galaxies or CSNG  (\S~\ref{subsec:control_sample}) to match the quasar sample. 
The CSNG serves as a control group. 
GAMA database ancillary products include estimates of SFRs for the comparison galaxies in the control group, as described in \S~\ref{subsec:sf_normals}, and we use those in this work.

\subsection{Low-redshift quasar sample}
\label{subsec:quasar_sample}

The low-redshift quasar sample is comprised of all known bright Type I AGN located within the equatorial volumes covered by the GAMA survey, and is taken from \cite{Wethers_2022}.
The quasar catalog \cite{Wethers_2022} is made up of 205 low-\textit{z} Type I quasars.
We briefly summarize how this sample of quasars was constructed: a full description is given in \cite{Wethers_2022}.

\cite{Wethers_2022} catalog of quasars was drawn from version 4 of the Large Quasar Astrometric Catalog (LQAC-4; \citealt{Gattano_2018}).
LQAC-4 is the most homogeneous and complete quasar catalog to date, cross-matching 12 independent quasar surveys, alongside the \cite{Veron_2010} quasar catalog.
LQAC-4 provides ubvgrizJK-band photometry, radio fluxes (at 1.4 GHz, 2.3 GHz, 5.0 GHz, 8.4 GHz, 24 GHz), and spectroscopic redshifts. 

The following selection constraints were applied by \cite{Wethers_2022} to the objects in the LQAC-4 catalog:
\begin{itemize}
    \item The quasars had to be within the three equatorial areas of the GAMA survey ({\fontfamily{qcr}\selectfont MagPhys} Data Management Unit, DMU).
    \item The redshift range was set to $0.1<z<0.35$, a redshift interval corresponding to a timescale of around 1~Gyr. 
    At this redshift range, the GAMA survey probes the largest volume with the highest completeness. 
    We note that within this relatively short time frame any possible evolutionary effects are likely to be less relevant. 
    \item The objects had to be brighter than $r=19.8$. 
    This selection criterion for apparent brightness was also imposed, since GAMA is complete beyond 99\% at $r<19.8$ mag (Sloan Digital Sky Survey, SDSS, photometry).
\end{itemize}

After the above selection procedure, \cite{Wethers_2022} proceeded to positionally match the LQAC-4 quasars to a GAMA target within $5''$: 90\% of the LQAC-4 quasars were identified as GAMA targets. 
The final quasar sample of 205 objects is also available as a Vizier catalog which includes the GAMA ID, coordinates, and redshifts for each quasar \citep{Wethers_2024Cat}.

All of these quasars also fulfill the criteria for inclusion in the SDSS DR12 catalog of Type I quasars \citep{Paris_2018}. The environment around this sample of quasars has been studied at different scales \citep{Wethers_2022,Stone_2023}.
Crucially, by selecting bright quasars within the GAMA equatorial footprint, we were able to utilize the full GAMA spectroscopic survey of several hundred thousand galaxies within the same patch of the sky to construct the control sample of normal galaxies for comparison, as described in the section (\S~\ref{subsec:control_sample}).

\subsection{Quasar sample properties}
\label{subsec:properties}

Here we consider the luminosity distribution (\S~\ref{subsec:lum}) and spectral types of quasars (\S~\ref{subsec:class}) in our GAMA sample. 
We compare our sample to the quasars from large surveys in the literature. 
This comparison confirms that our quasar sample (which is constrained to GAMA equatorial volume) consists of typical quasars. 
(The star formation properties of the quasar sample are discussed in the next section, \S~\ref{sec:star_properties}.)

\subsubsection{Luminosity of quasar sample}
\label{subsec:lum}

\begin{figure}[hbt]
\centering
    \subfigure[]{\includegraphics[width=\columnwidth]{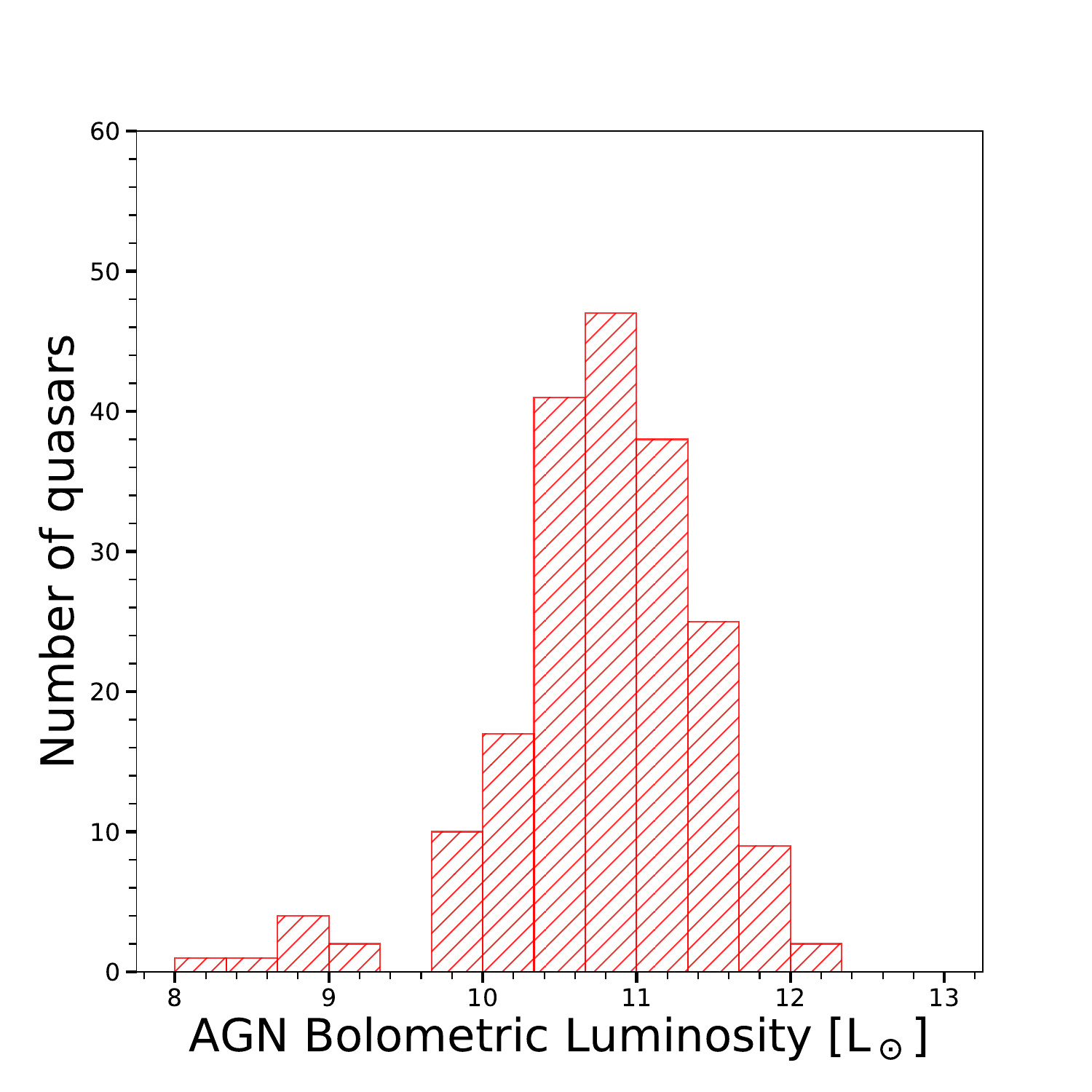}}\\
    \subfigure[]{\includegraphics[width=\columnwidth]{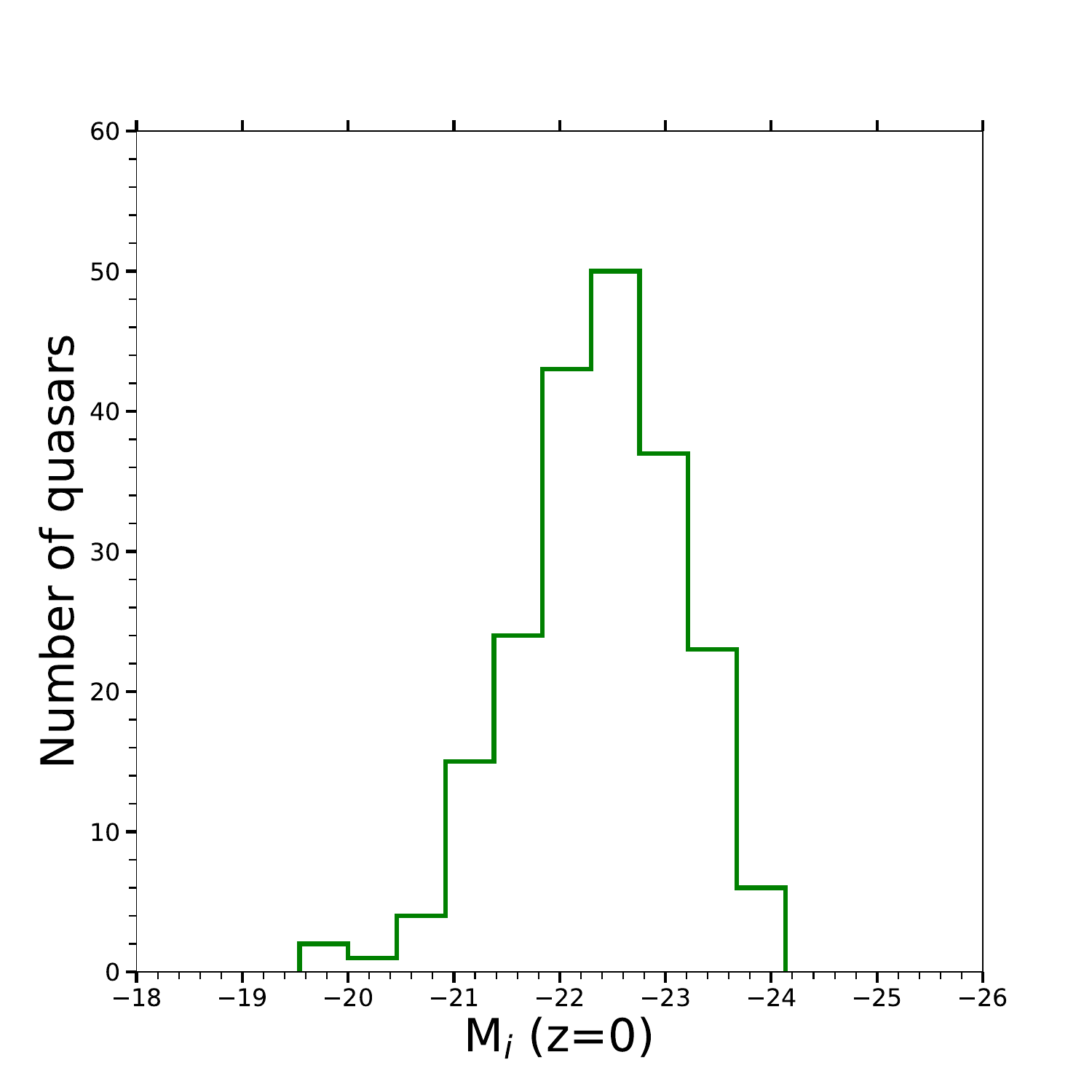}}
\caption{
(a) Distribution of bolometric luminosities for quasars derived from \textsc{cigale} fits. 
This excludes 8 quasars (3.9 \% of quasar sample) where \textsc{cigale} did not return a quasar luminosity.
(b) Distribution of absolute SDSS $i$ band magnitude ($k$-corrected and extinction corrected based on GAMA survey data). 
}
\label{fig:lum}
\end{figure}

The AGN bolometric luminosity and the distribution of the absolute luminosities in the $i$ band are shown in Fig.~\ref{fig:lum}.
The bolometric luminosity is characteristic of low to intermediate luminosity objects.
This is not surprising, as powerful quasars will be rare within these relatively small volumes. 

\subsubsection{Classification of quasars}
\label{subsec:class}

We classified \textit{by eye} the optical spectra of quasars using the spectra in the GAMA archive. We used three broad categories: quasars with at least one broad line (Seyfert 1 or S1), quasars with narrow lines (Seyfert 2 or S2), and systems with low-ionization nuclear emission-line regions (LINERs), following the classification scheme of the \cite{Veron_2010} catalog. 
The results are presented in Table~\ref{tab:spectral_types}.
We further compared our visual classification with \cite{Veron_2010} catalog classifications, and found them mostly consistent with each other.

\begin{deluxetable}{lcc}
\tablecaption{Categories of quasars based on spectral type in this work versus the literature quasar catalog \cite{Veron_2010}. \label{tab:spectral_types}}
\tablehead{
\colhead{Category} & \colhead{This Work} & \colhead{\cite{Veron_2010}}\\
\colhead{}& \colhead{$N=205$} & \colhead{$N=9586$}
}
\startdata
S1          &    192 (94 \%)         & 9353 (97 \%) \\%[2pt]
S2          &    13 (6 \%)           &      0       \\%[2pt]
S3 (LINER)  &    0                   & 233 (3 \%)
\enddata
\tablecomments{
Column (1): Quasar classification category. 
Column (2): Category fractions of the full quasar sample used in this work. 
Column (3): Category fractions of quasars from literature.}
\end{deluxetable}

If no classification was given in \cite{Veron_2010}, then, when possible, for these rare cases we adopted classifications from more recent catalogs in the Vizier Service (e.g., \citealt{Rakshit_2017}).
A small number of quasars, however, did not have any spectral classification in the existing catalogs and we used the visual classifications as the final determination. 

Most of our galaxies are Seyfert 1 with typical broad emission lines. 
The mean redshift of the quasar sample is $z=0.224\pm0.072$, while the mean absolute magnitude is $M_i=-22.38\pm0.77$. 
Compared to the previous low-redshift study of Type I quasars by \cite{Falomo_2014}, our sample of quasars probes also lower brightness AGN, due to the different cuts imposed by quasar catalogs (e.g., SDSS quasar catalog) available at the time. 
Our sample is essentially consistent with that of \cite{Veron_2010} as shown in Table~\ref{tab:spectral_types} and appears to be representative of the average population of AGN within this redshift range ($0.1<z<0.35$).

\subsection{Comparison sample of normal galaxies}
\label{subsec:control_sample}

It is important to compare any results from our quasar sample with a control sample of normal galaxies, in order to ascertain that the observed characteristics are indeed unique to quasar host galaxies \citep[as discussed previously in e.g.,][]{Karhunen_2014}. 
The GAMA spectroscopic survey database is whence the normal galaxies for CSNG are selected, because in addition to the robust redshifts, the GAMA archive provides estimates of physical properties \citep{Driver_2022}.
Thus, the match between the active and normal galaxies is robust, since both the redshift and the physical properties are matched between the two samples.

\begin{figure}
    \centering
    \includegraphics[width=\columnwidth]{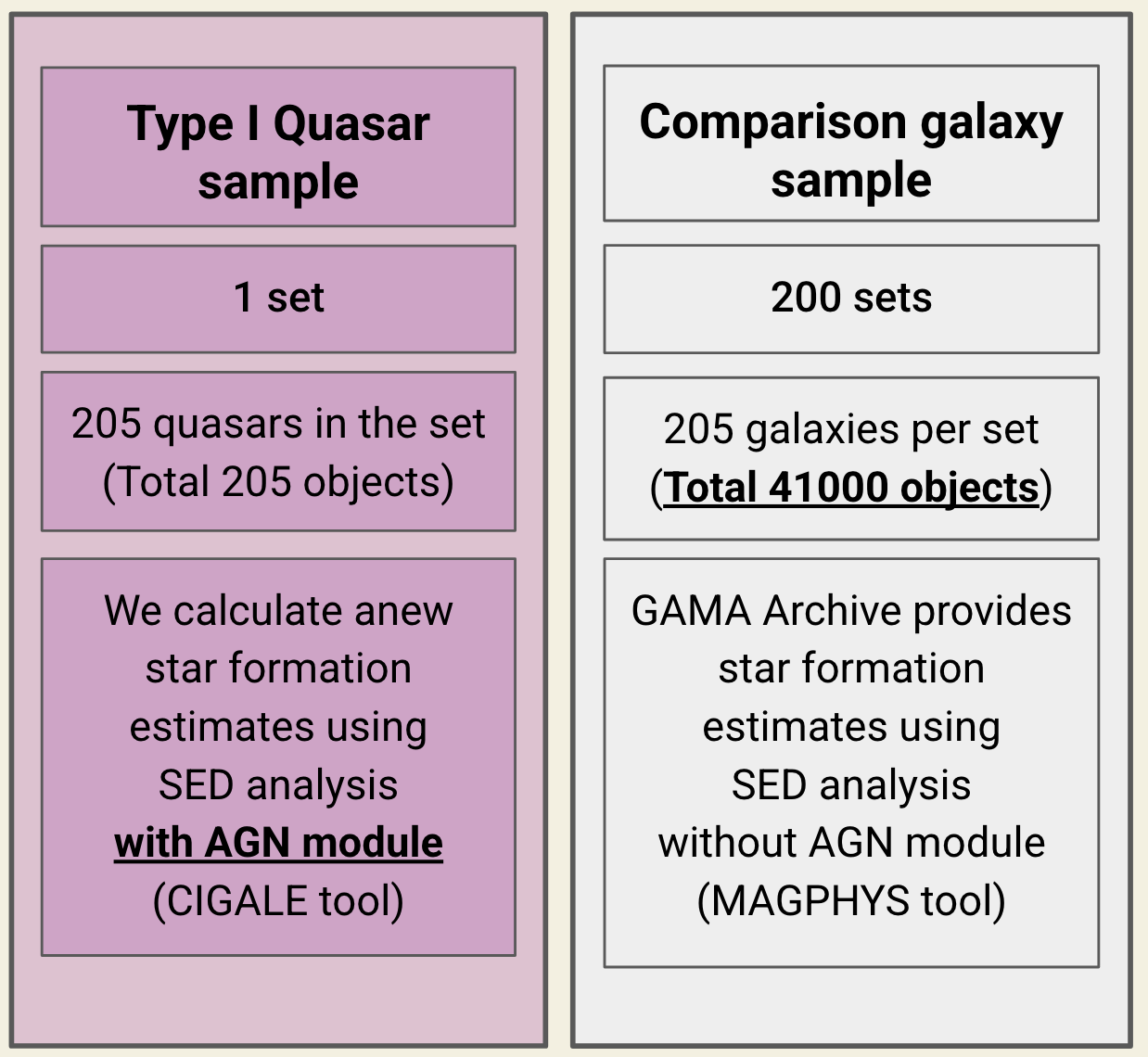}
    \caption{Our quasar sample consists of 205 objects. 
    Our CSNG is constructed using the GAMA archival data and consists of 200 sets, where each set has 205 normal galaxies. 
    The CSNG is from the same volume as the quasar sample. 
    The redshift and stellar mass properties are matched to the quasar sample as well (Fig.~\ref{fig:match}).
    The quasar sample is obtained from the \cite{Wethers_2022} and the CSNG is obtained from \cite{Stone_2023}.}
    \label{fig:samples}
\end{figure}

\begin{figure}[hbt]
    \centering
    \subfigure[]{\includegraphics[width=0.45\textwidth]{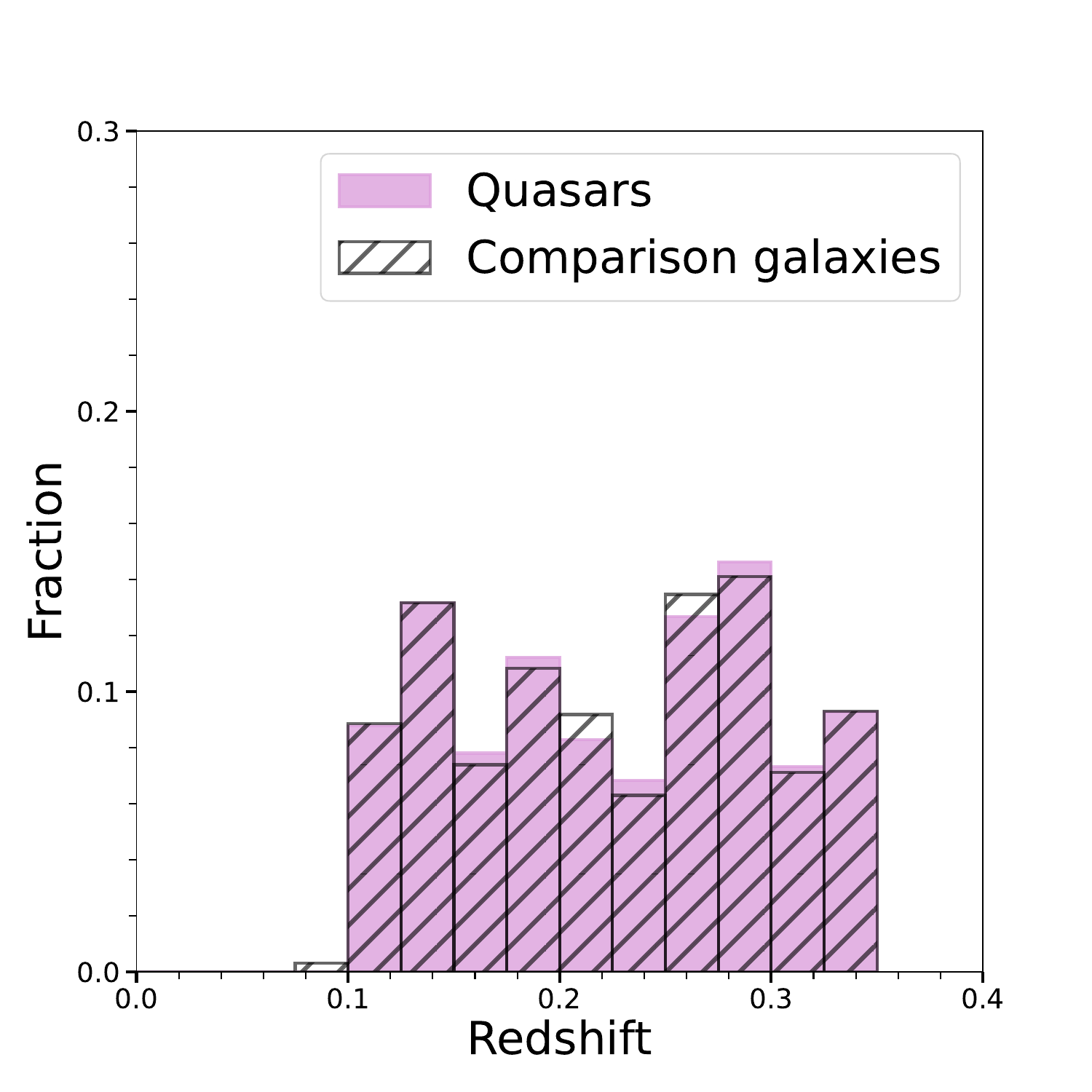}}\\
    \subfigure[]{\includegraphics[width=0.45\textwidth]{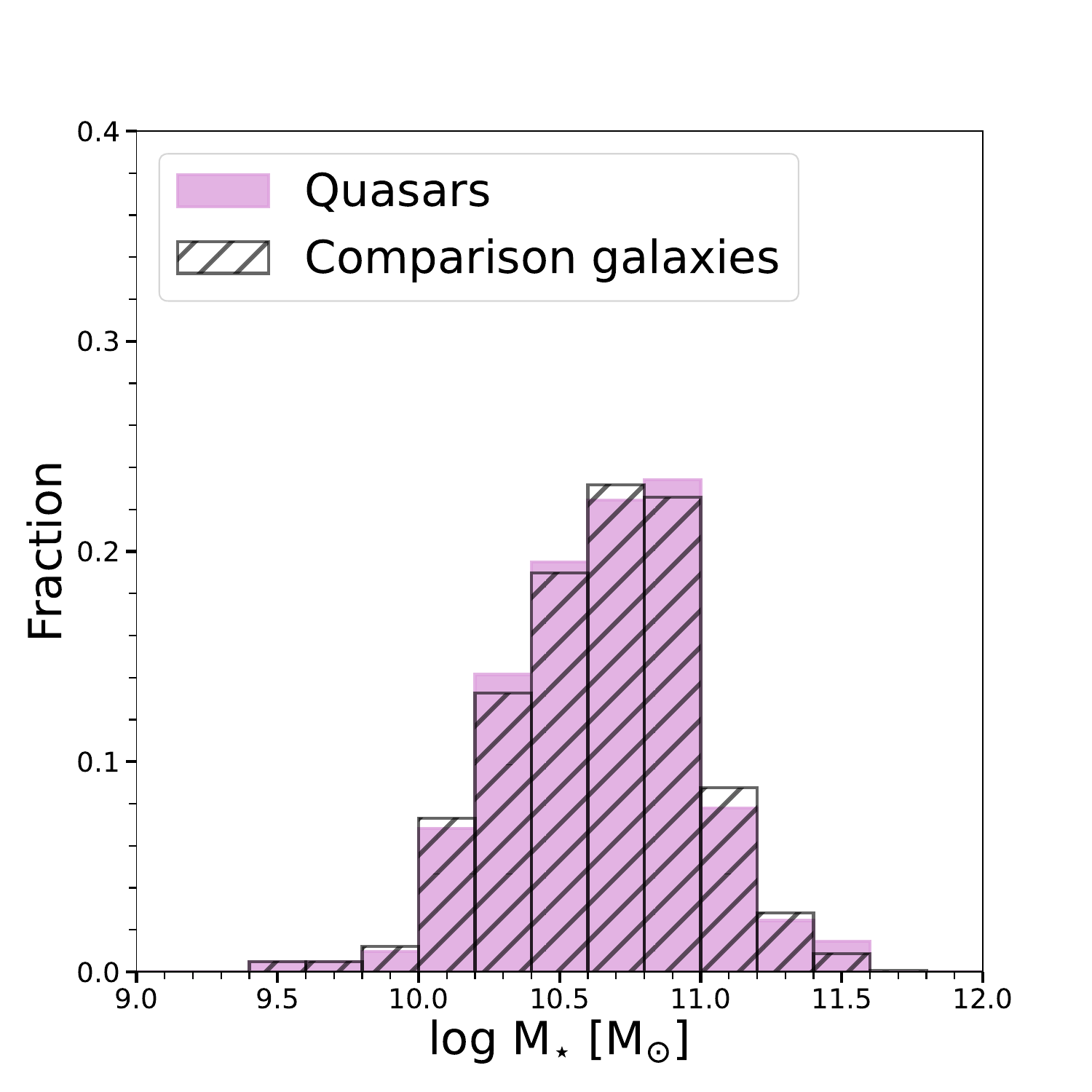}}
    \caption{(a) Redshift and (b) stellar mass distributions of quasars and comparison galaxies (reproduction of Fig.~1 from \citealt{Stone_2023}).
    The two-sample Kolmogorov-Smirnov test ($p-value>0.99$) shows that there is no significant difference statistically between the matched normal galaxies sample and the quasar sample (same result with the Anderson-Darling test).}
    \label{fig:match}
\end{figure}

The CSNG for our quasar sample is obtained from \cite{Stone_2023} and consists of 200 sets of normal galaxies (Fig.~\ref{fig:samples}). 
All galaxies were extracted from the GAMA spectroscopic survey ({\fontfamily{qcr}\selectfont MagPhys} DMU). 
Each set of normal galaxies has:
\begin{itemize}
    \item 205 normal galaxies, as in our quasar sample,
    \item is within the same volume as the quasar sample, and
    \item is matched in redshift and stellar mass to the quasar sample (Fig.~\ref{fig:match}).
\end{itemize}
Note that the CSNG does not have restrictions on the star formation rates, and control galaxies range from star-forming to objects with low star formation activity (similar to \citep{Bettoni_2015}).

The matching method used by \cite{Stone_2023} to build these comparison sets of normal galaxies was based on a \textit{Monte Carlo} realization of samples technique, and this method was first presented in \cite{Wethers_2022}. 
Briefly, for each quasar, a randomly selected galaxy having the same redshift within $\Delta z = 0.01$ and the same stellar mass within 0.1 dex is selected as an normal counterpart.
This process is repeated for all quasars, until a comparison set of 205 GAMA non-AGN galaxies is defined. 
In the same way, 200 such comparison sets were created. 
As shown in Fig.~\ref{fig:match}, this results in closely matched samples of galaxies and quasars, as checked both by the Kolmogorov-Smirnov and the Anderson-Darling statistical tests.

\subsection{GAMA archive star formation properties for CSNG}
\label{subsec:sf_normals}

The star formation properties for normal galaxies were retrieved from the archives of the GAMA database.
GAMA derives estimates for the stellar population properties (such as SFRs and SFHs) by fitting the SEDs of galaxies with the \textsc{magphys} tool \citep{daCunha_2008, daCunha_2011}. 
The SEDs used by GAMA are based on the 21$-$band fluxes for the equatorial survey regions (Table~\ref{tab:bands}), covering wavelengths from far-UV to far-infrared ({\fontfamily{qcr}\selectfont LambdarPhotometry} DMU, \citealt{Driver_2016,Wright_2016}).

\begin{deluxetable*}{ll}
\tablecaption{GAMA spectroscopic survey photometry bands. \label{tab:bands} }     
\tablehead{
\colhead{Bands} & \colhead{Survey}
}
\startdata
FUV, NUV & \textit{Galaxy Evolution Explorer} (\textit{GALEX}) \citep{Martin_2005} \\
$u$, $g$, $r$, $i$, $z$ & SDSS Data Release (DR) 7, \citealt{Abazajian_2009}\\
Z, Y, J, H, Ks & Visible and Infrared Survey Telescope for Astronomy (VISTA) Kilo-Degree Infrared Galaxy Survey\\
& (VIKING), \citealt{Edge_2013}\\
W1-W4, $\sim$2--30~$\mu$m & \textit{Wide-Field Infrared Survey Explorer} (\textit{WISE}) All-Sky DR \citep{Wright_2010}\\
100~$\mu$m, 160~$\mu$m & \textit{Herschel} Astrophysical Terahertz Large Area Survey \citep{Eales_2010} PACS\\
250~$\mu$m, 350~$\mu$m, 500~$\mu$m & \textit{Herschel} SPIRE
\enddata
\tablecomments{
Column (1): GAMA {\fontfamily{qcr}\selectfont LambdarPhotometry} DMU Photometry bands. 
Column (2): Source surveys.
}
\end{deluxetable*}

The SED-based estimates are drawn from within the {\fontfamily{qcr}\selectfont MagPhys} DMU of the GAMA survey \citep{Driver_2022}. 
We extracted the following properties from the {\fontfamily{qcr}\selectfont MagPhys} DMU for CSNG: stellar masses, SFRs and SFRs averaged over the past 10, 100, 1000 and 2000~Myr, plus the estimates for the averaged SFR for older stellar populations.
While there are estimates in GAMA archive also for our quasar sample, we perform a separate analysis to estimate the star formation properties for quasar hosts, in order to include the most up-to-date SED models which account for the presence and contribution of the AGN, as described in the next section (\S~\ref{sec:star_properties}).
Additionally we perform similar \textsc{cigale} fits for comparison on a representative subsample of CSNG, described in (\S~\ref{sec:tests}), to make sure that the comparison between the quasar hosts and normal galaxies is homogeneous.

\section{SFR and SFH estimates for quasar hosts}
\label{sec:star_properties}

In this section, we present the necessity and justification for rederiving the estimates for star formation properties for the quasar sample with the \textsc{cigale} SED tool (\S~\ref{sec:cigale}). 
Then we present the details on how \textsc{cigale} SED fits were setup for the quasar hosts (\S~\ref{sec:cigalefits}).
We describe few comparison tests between \textsc{cigale} and \textsc{magphys} (\S~\ref{sec:tests}).
In particular, an additional check was performed for a single set of normal galaxies from the control group.
We further introduce a statistical correction between the two SFR estimators.
For completeness, a brief discussion is included to describe the underlying assumptions such as initial mass function (IMF) for both codes.

\subsection{\textsc{cigale} AGN module}
\label{sec:cigale}

Accounting for the AGN contribution is necessary in assessing the quasar host star formation properties; lack of such consideration results in overestimation of the parameters.
Thus, we adopt the \textsc{cigale} SED fitting code to analyze all objects in our quasar sample, because \textsc{cigale} allows the inclusion of an AGN component in SED fits \citep{Burgarella_2005,Noll_2009,Boquien_2019,Yang_2022}.

To account for AGN contribution, we chose the most up-to-date AGN model, the SKIRTOR module.
The SKIRTOR module has a realistic two-phase torus model.
Additionally, SKIRTOR AGN module covers the UV-IR range which matches the GAMA photometry interval.

For quasar host galaxies in our sample, it was not possible to use the star formation estimates from the available GAMA archive because the GAMA SED fits did not include the AGN contribution.
In addition, it was not possible to use the \textsc{magphys} SED fitting code, as in the GAMA archive, because that code did not have models for the AGN contribution. 
The \textsc{magphys} team started developing this capability recently (da Cunha, Juneau et al., in prep.).
We discuss the differences between the codes later in this section.

The separation of AGN from galaxy light has been treated and tested by the \textsc{cigale} team \citep{Boquien_2019, Yang_2019, Yang_2022}, including for the case of the SDSS Type I quasars, i.e., the same category of quasars as in this work.
Numerous later studies used \textsc{cigale} to tackle the AGN contribution in SED analysis and confirm that this method is effective and reliable both at low and at high redshifts \citep[e.g.,][]{Circosta_2018,Mount_2022b,Mount_2022,Georgan_2023}. 
This is generally true even for the most luminous quasars, although these are not generally present within the GAMA volume (see \S~\ref{subsec:properties}).

We stress that for normal galaxies the estimates of SFR and similar quantities are published in GAMA archives, as described in \S~\ref{subsec:sf_normals}.
As there is no need to include the AGN component in SED fits for normal galaxies, the fits provided by GAMA are satisfactory; it is not necessary to rederive them again with \textsc{cigale} (which would be a computationally intensive task as the comparison sample consists of more than 40000 objects).
Nevertheless, we did run comparison tests with \textsc{cigale}for a representative subsample of control galaxies, described in \S~\ref{sec:tests}.
Furthermore, as presented in that same section, extensive comparative analysis of both codes by \cite{Hunt_2019} asserts that the resultant estimates from both codes are similar and equally reliable.

\subsection{\textsc{cigale} parameters for the quasar SED fits}
\label{sec:cigalefits}

We use the same multiband data from the GAMA Lambdar Catalog for building \textsc{cigale} fits. 
We used version 2022.1 of \textsc{cigale}, which is \textsc{python}-based. 

\textsc{cigale} uses a series of modules for SFH, single stellar population (SSP), and other contributions to build the final SED of a galaxy \citep{Boquien_2019}. 
The chosen modules, parameters, and their values for the quasar hosts with AGN component are given in Table~\ref{table:parameters}. 
Neither X-ray nor radio-data are present in GAMA survey archive, so these modules are not included.

\begin{deluxetable*}{ll}
\tablecaption{Model parameters for SED fitting with \textsc{cigale}. 
The AGN component (SKIRTOR) was included only for quasar fits.
For further details on the parameters, see the \textsc{cigale} manual and \citealt{Boquien_2019, Yang_2022}. \label{table:parameters}}
\tablehead{
\colhead{Module and Parameter (Symbol)} & \colhead{References and Values}}
\startdata
\textbf{SFH} & \textit{Delayed SFH with an optional recent burst } \\
Age of the main population (Myr) & 3000, 5000, 7000, 8000, 9000, 10000\\
e-folding time of the main population (Myr), $\tau_{main}$ & 1000, 3000, 5000, 8000, 10000 \\
Age of the recent burst (Myr) & 50, 20000 (continuous SFH) \\
e-folding time of the burst (Myr) & 50, 500 \\
Burst mass fraction & 0.0, 0.1 \\[2pt]
\textbf{SSP} & \textit{\cite{Bruzual_2003}} \\
IMF & \cite{Chabrier_2003} \\
Metallicity (\textit{Z}) & 0.02 (Solar)\\[2pt]
\textbf{Galactic dust attenuation} & \textit{Modified Calzetti (2000) attenuation law} \\
Color excess of nebular lines & 0.1, 0.2, 0.3, 0.4, 0.5, 0.6 \\
Reduction factor & 0.44 \\
Slope of the power law & -0.2, 0.0 \\[2pt]
\textbf{Galactic dust emission} & \textit{\cite{Dale_2014}} \\
Alpha slope & 2.0 \\[2pt]
\textbf{AGN (UV-to-IR) } & \textit{SKIRTOR} \citep{Stalevski_2012,Stalevski_2016} \\
Viewing Angle & 30\textdegree \,(Type I) \\
Delta & -1,-0.9,-0.8,-0.7,-0.6,-0.5,-0.4,-0.3,-0.2,-0.1,0.1,0.6 \\
AGN fraction & 0.0,0.01,0.1,0.2,0.3,0.4,0.5,0.6,0.7,0.8,0.9,0.99 \\
$E(B-V)$ (magnitudes) & 0., 0.05, 0.1, 0.2, 0.4, 0.6 \\
Extinction law of polar dust & Small Magellanic Cloud
\enddata
\tablecomments{
Column (1): List of parameters in each module of the \textsc{cigale} SED tool used in this work. 
Column (2): Parameter values adopted in our SED analysis with \textsc{cigale} tool.}
\end{deluxetable*}

We assume the \textit{delayed} SFH scenario with a star formation burst \citep{Boquien_2019, Yang_2022}. 
The library of SSP templates is from \cite{Bruzual_2003}, allowing for the construction of the stellar emission part of the SED from the galaxy host. 
We use the Chabrier IMF \citep{Chabrier_2003} and assume solar metallicity ($Z=0.02$) following \cite{Ciesla_2015}. 
The stellar emission is attenuated by applying the \cite{Calzetti_2000} dust extinction law.

The contribution due to dust heating by the stellar component is modeled by adopting the \cite{Dale_2014} dust templates with parameter values as in \cite{Ciesla_2015}. 
To include the non-stellar emission contribution from the AGN, we use the newer AGN templates --- the SKIRTOR templates \citep{Stalevski_2012, Stalevski_2016}, but only for Type I quasars, since our sample only contains Type I AGN.  
For the input parameters, we use the suggestions from \cite{Yang_2022} for the SDSS quasar case.

For 8 objects, SED fit had zero AGN luminosity; however, this does not affect the conclusions of this work.
We do not force the SED fit to choose quasar models, so in these 8 cases, the best fit found in the library was not the SED of quasar.
We chose not to add additional parameters, because it is not possible to get a perfect fit for every single quasar in the sample; we employ a single set of parameters to fit SEDs for the full quasar sample.
Enlarging the parameter space may result in overfitting and in different sets of misfits. 
The parameter grid size is a balance between fitting well as many quasars as possible in the sample and avoiding overfitting and high computational pressure.
The quality of fits is primarily limited by the uncertainties in the input photometry and by the content of \textsc{CIGALE} library of SEDs.

\subsection{Comparing our \textsc{cigale} estimates to GAMA data}
\label{sec:tests}

We perform two comparison tests to check how the estimates based on the \textsc{cigale} fits compare to the archival data from the GAMA spectroscopic survey (Table~\ref{tab:comparison_tests}).
We perform both a test for the quasar sample (test A) and a test for the CSNG (test B).

\begin{deluxetable*}{llllr}
\tablecaption{Comparison tests of star formation properties between the derivations with \textsc{cigale} fits in this work and GAMA archival data. \label{tab:comparison_tests}}
\tablehead{\colhead{Test ID} & \colhead{Population} & \colhead{Comparison description} & \colhead{\textsc{cigale} AGN module} & \colhead{Figure reference}}
\startdata
A & quasar sample & \textsc{cigale} fits to GAMA data & included & \ref{fig:comparisons_quasars_withAGN}, \ref{fig:comparisons_SFR_averages_quasars} \\%
B & a single set of galaxies from CSNG & \textsc{cigale} fits to GAMA data & not included & \ref{fig:comparisons_gals_withoutAGN}, \ref{fig:comparisons_SFR_averages}
\enddata
\tablecomments{
Column (1): Test ID.
Column (2): Population compared.
Column (3): AGN module inclusion in \textsc{cigale} fits. (Note that in all cases GAMA archival data does not include AGN contribution to the SED fits.)
Column (4): Figures with test outcome.
}
\end{deluxetable*}

\subsubsection{Test A: Check the effect of AGN contribution to the SED fits}
\begin{figure*}
    \centering
    \subfigure[]{\includegraphics[width=0.6\textwidth]{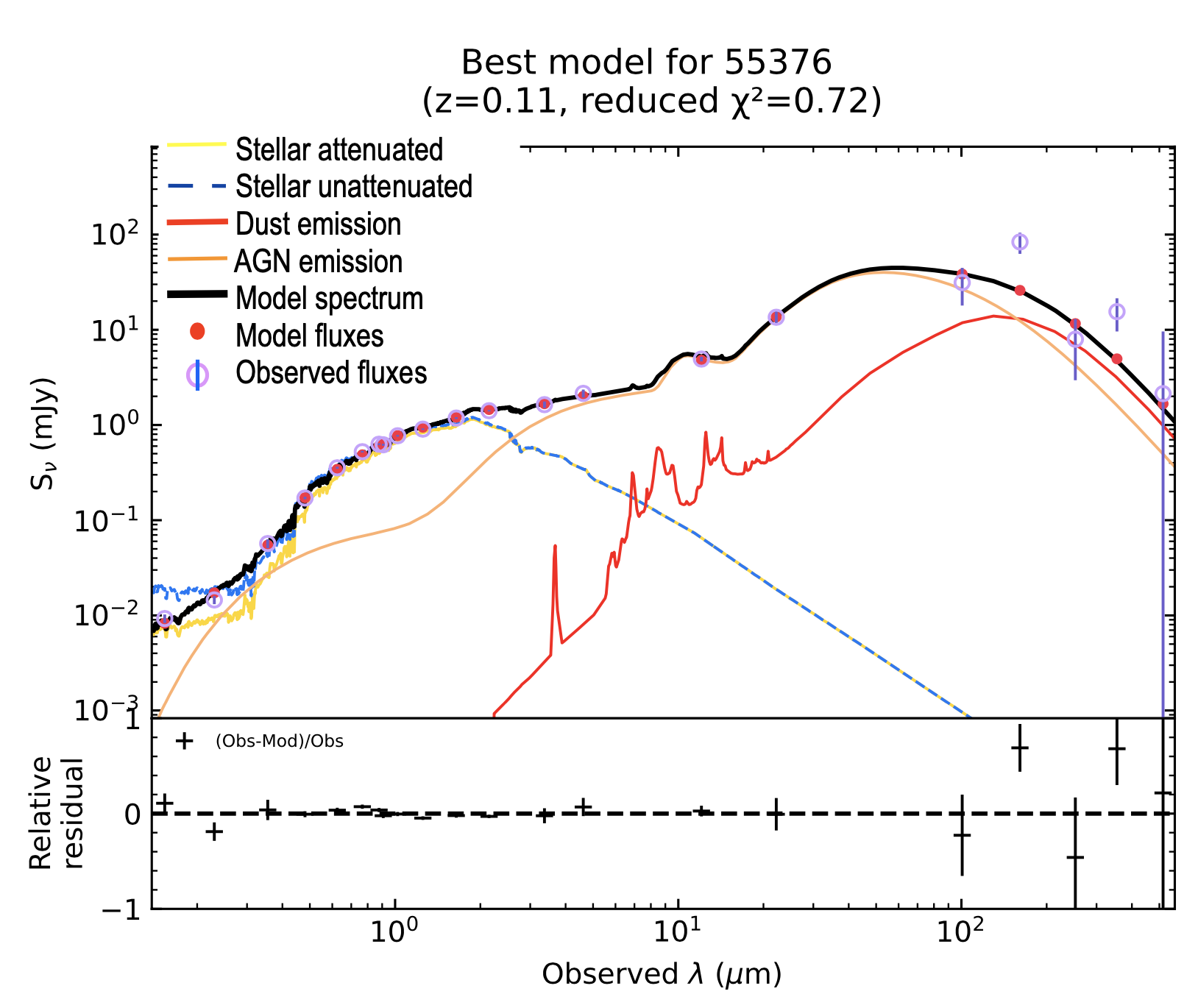}}\\
    \subfigure[]{\includegraphics[width=0.6\textwidth]{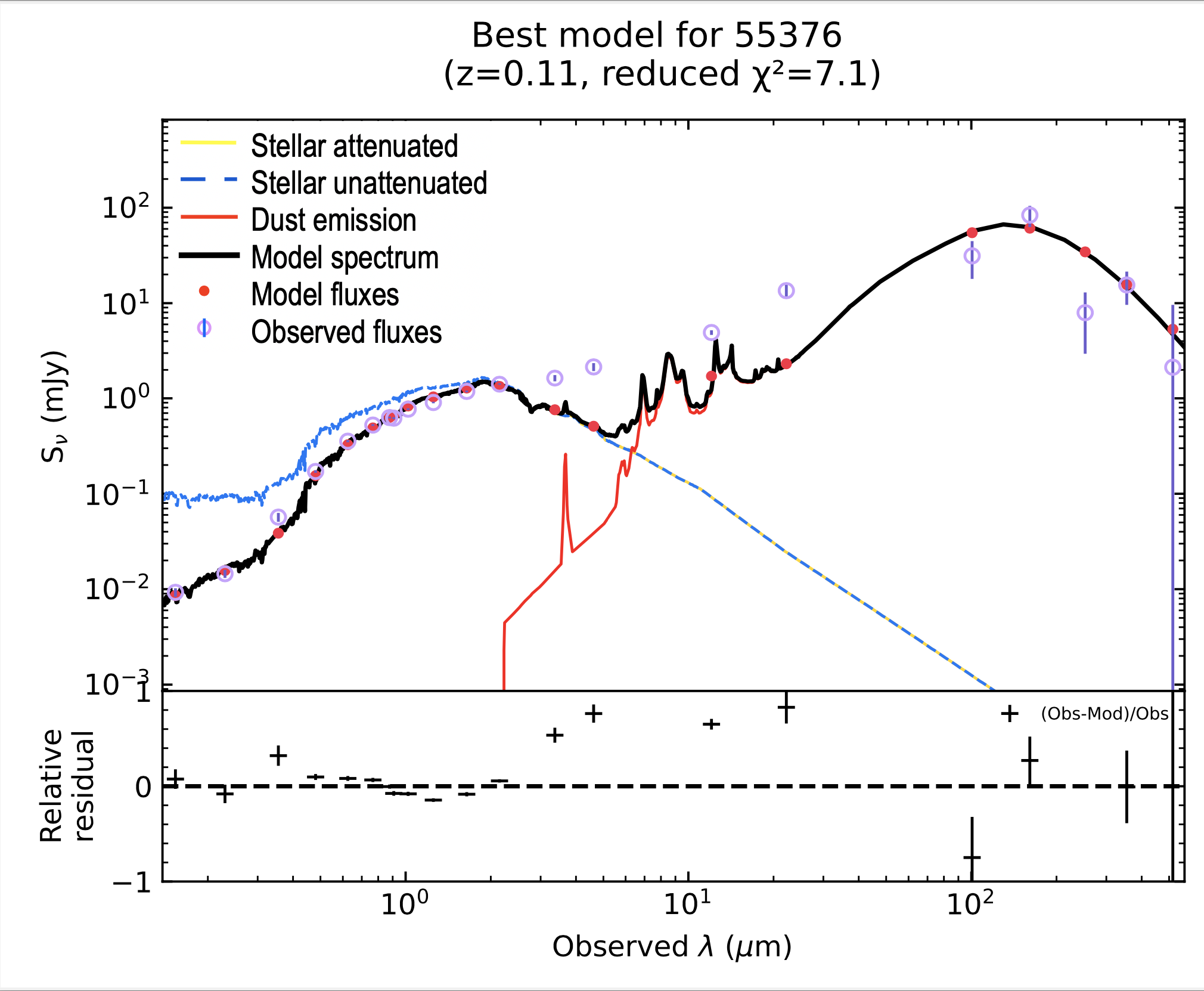}}
    \caption{SED analysis result with \textsc{cigale}for one quasar in our sample.
    (a) Top panel shows the result which included AGN into consideration. 
    AGN emission is represented by the solid orange line.
    (b) Bottom panel shows the best model without inclusion of the AGN module.
    Note the reduced $\chi^{2}$ value is much lower when AGN is considered, indicating a better fit.}
    \label{fig:example}
\end{figure*}

\begin{figure*}[hbt]
    \centering
    \subfigure[]{\includegraphics[width=0.45\textwidth]{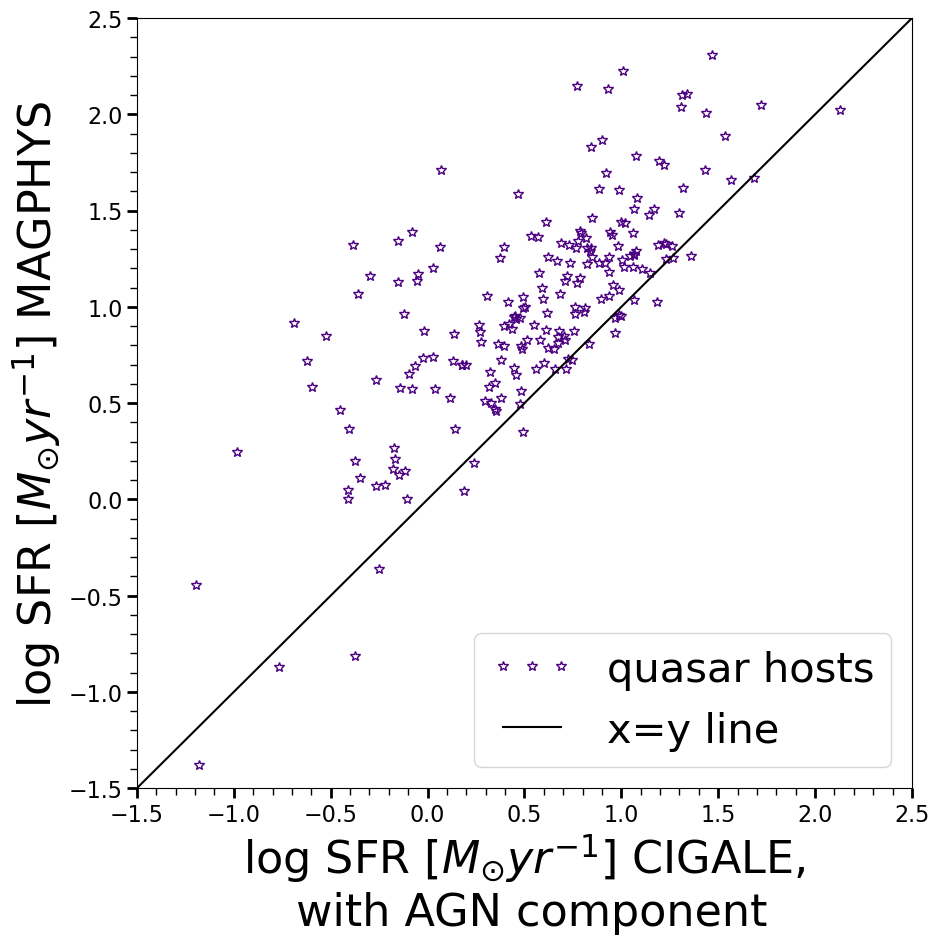}}  
    \subfigure[]{\includegraphics[width=0.45\textwidth]{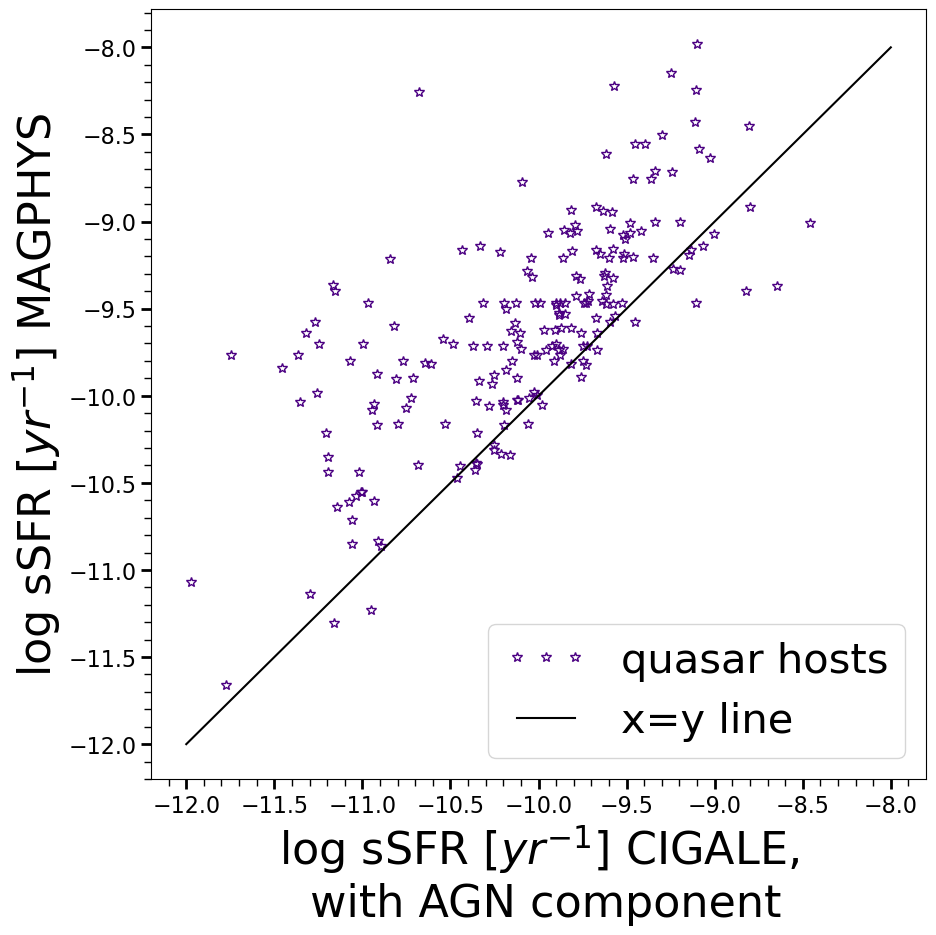}}\\
    \subfigure[]{\includegraphics[width=0.45\textwidth]{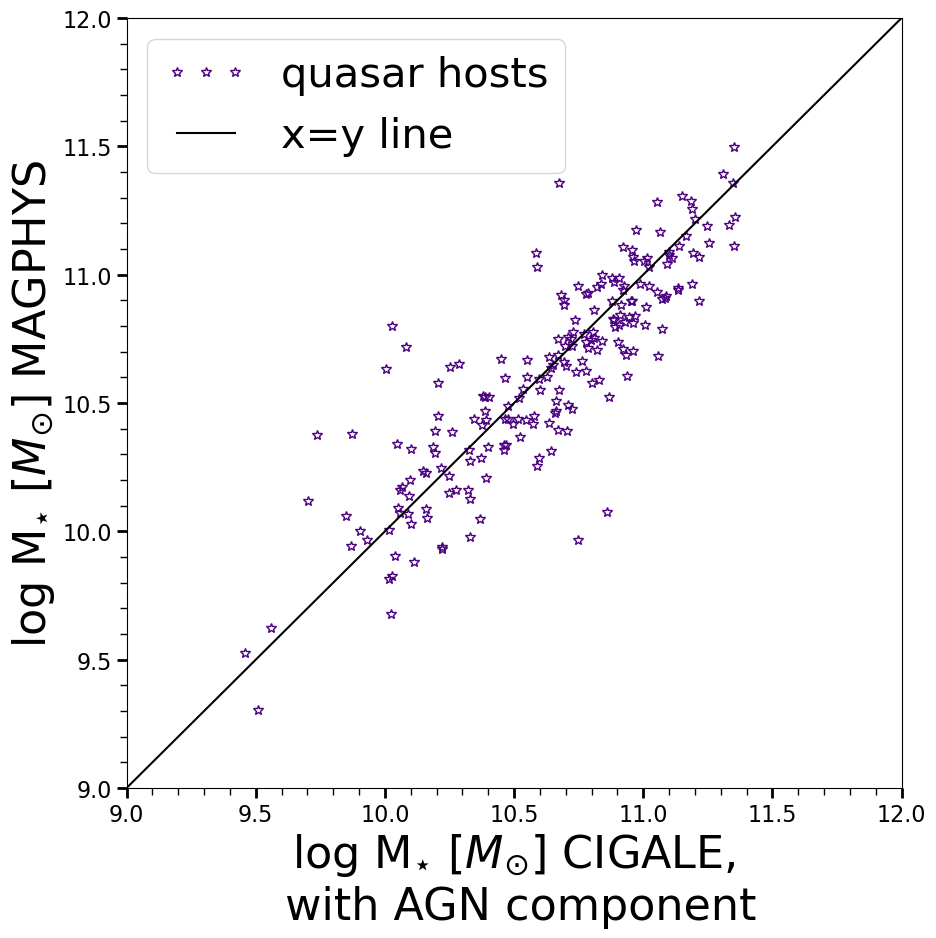}}
\caption{Results for comparison test A. 
Comparison between the GAMA archival data (\textsc{magphys} SED fitting results with no AGN component) and our \textsc{cigale} SED fitting results (including an AGN component) for (a) SFR, (b) sSFR, and (c) stellar mass for the sample of quasar host galaxies. 
The SFR and sSFR estimates differ significantly when an AGN contribution is included in the SED analysis, prompting us to derive anew the star formation property estimates for the quasar sample.}
\label{fig:comparisons_quasars_withAGN}
\end{figure*}

In test A, we compare our \textsc{cigale} results with AGN component to the archival GAMA data for quasar hosts which did not include the AGN contribution.
GAMA archive is based on \textsc{magphys} SED fits, as noted previously.
We show one example of quasar in our sample analyzed both with AGN and without AGN in Fig.~\ref{fig:example}.
Comparisons for the SFR, specific SFR (sSFR), and stellar mass are given in Fig.~\ref{fig:comparisons_quasars_withAGN}. 
Not surprisingly, \textsc{magphys} tends to overestimate the SFRs (ignoring the AGN component) while the derived stellar masses are comparable.
Thus, this test result shows that it is important to consider the AGN component when performing SED-based analysis to derive properties for quasar hosts.

\subsubsection{Test B: Check the agreement between two codes for CSNG}
\begin{figure*}[hbt]
    \centering
    \subfigure[]{\includegraphics[width=0.5\textwidth]{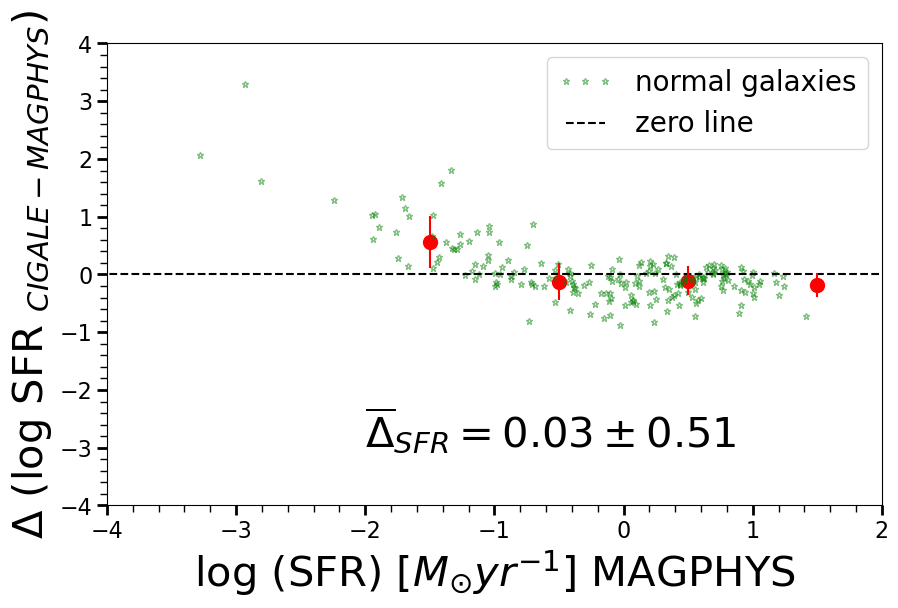}}\\
    \subfigure[]{\includegraphics[width=0.5\textwidth]{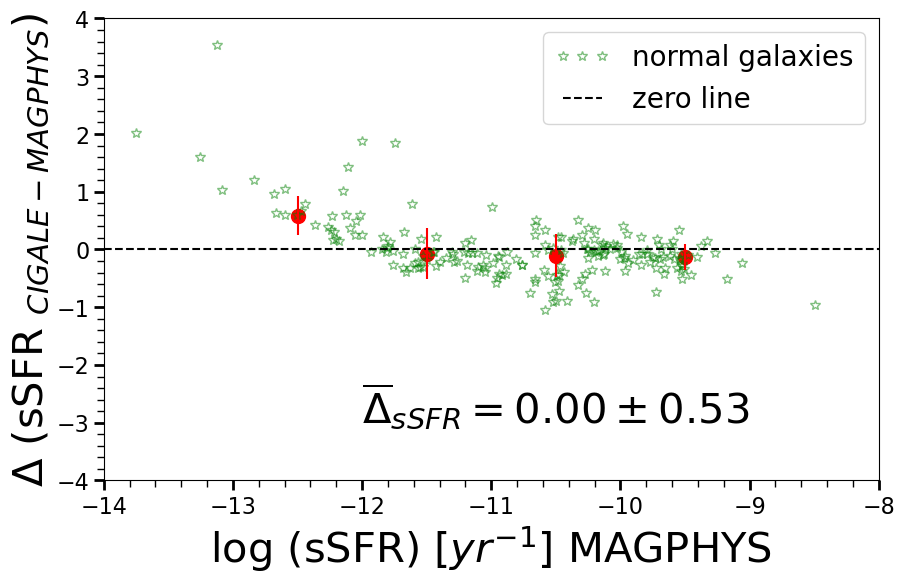}}\\
    \subfigure[]{\includegraphics[width=0.5\textwidth]{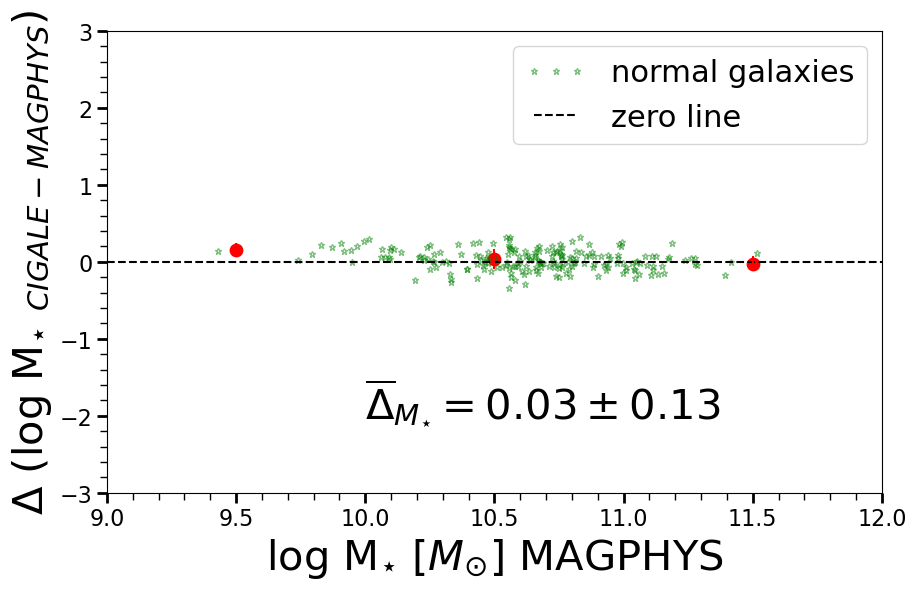}}
\caption{Results for comparison test B. 
Differences ($\Delta$) comparing the \textsc{magphys} and \textsc{cigale} SED fitting results (without AGN component) for (a) SFR, (b) sSFR, and (c) stellar mass for normal galaxies.
The red filled circles represent the median log difference in each 1 dex bin of SFR between the two estimators, excluding the bins with small number statistics. 
Error bars reflect the standard deviation in that bin. 
We remark that in all half dex intervals the only difference between \textsc{cigale} and \textsc{magphys} is a small (on the order of 0.01 dex) offset with no particular trend, except at very low SFR where essentially all models can yield different results based on small differences in input parameters.}
\label{fig:comparisons_gals_withoutAGN}
\end{figure*}

It is expected that for normal galaxies the estimates of properties should not depend on which SED tool is used, \textsc{cigale} or \textsc{magphys}.
Thus, next, in test B, we considered the normal galaxies and compared the results of \textsc{cigale} fits (with no AGN component) to GAMA archive data (based on \textsc{magphys}).
This test was done for \textit{one} CSNG set.
Rather than running \textsc{cigale}on thousands of GAMA galaxies, we selected a representative sub-sample, matching the quasar sample (same number of objects, same distribution in redshift and mass, same survey volume).
We perform Test B to make sure that the comparison between quasars and non-active galaxies is homogeneous.
The test result for SFR, sSFR, and stellar masses is given in Fig.~\ref{fig:comparisons_gals_withoutAGN}.

The SED fits with \textsc{cigale} for normal galaxies were setup identically to the quasar hosts, with the exception of AGN module; AGN component was not included.
Table~\ref{table:parameters} gives the list of all parameters for the \textsc{cigale} fits for normal galaxies as well.
The expectation in test B is that it does not matter which SED software is employed to estimate SFRs, so the expected output should result in similar results.
Indeed, this is what we observe, as shown in Fig.~\ref{fig:comparisons_gals_withoutAGN}.
For the vast majority of normal galaxies there is virtually no difference between the two estimates with two different codes.
Only for a few objects with extremely low SFR values, the differences between the two estimates deviate from zero.
However, both algorithms suffer from poor performance at this extreme regime in SFR, and it is not possible to say that one evaluates the extremely low SFR values more precisely.
As the number of objects in that regime is small, our results and conclusions are not affected.

\begin{figure*}[hbt]
    \centering
    \subfigure[]{\includegraphics[width=0.47\textwidth]{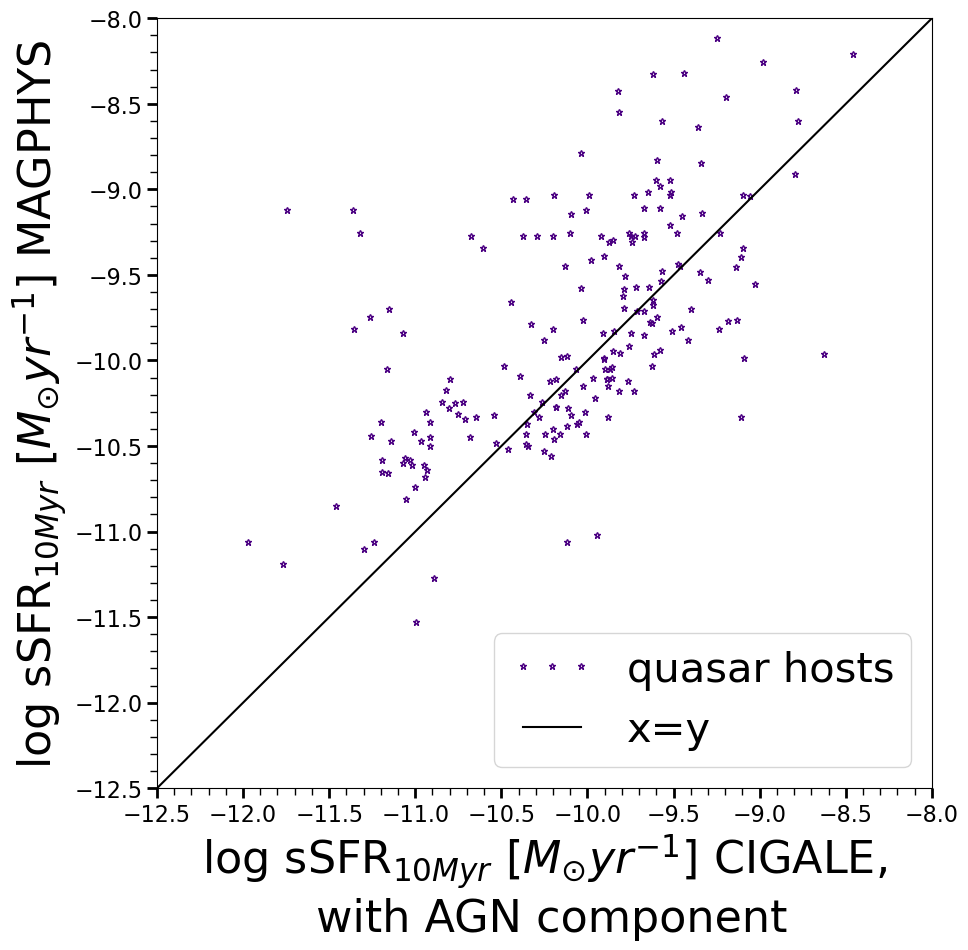}}
    \subfigure[]{\includegraphics[width=0.47\textwidth]{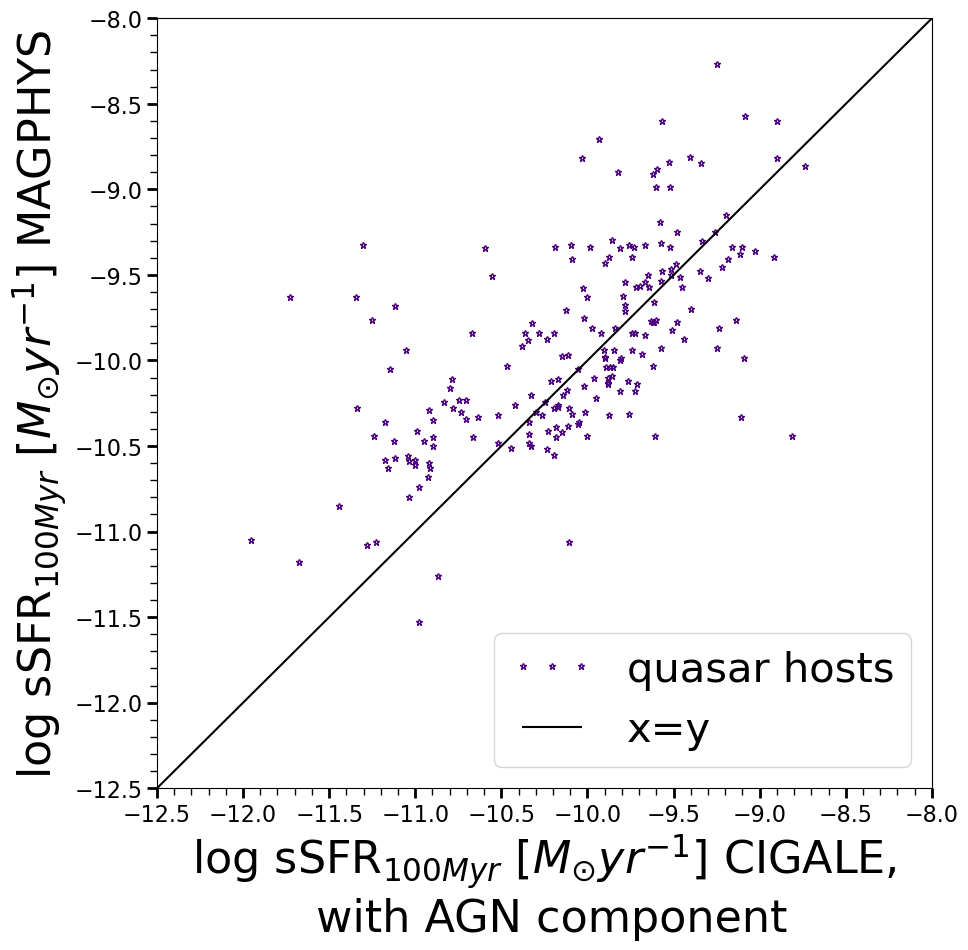}}
\caption{Comparison between the \textsc{magphys} (no AGN component) and \textsc{cigale} (with AGN component) SED fitting results for sSFR averaged over the past (a) 10 and (b) 100~Myr, for quasar hosts (test A).}
\label{fig:comparisons_SFR_averages_quasars}
\end{figure*}

\begin{figure*}[hbt]
    \centering
    \subfigure[]{\includegraphics[width=0.48\textwidth]{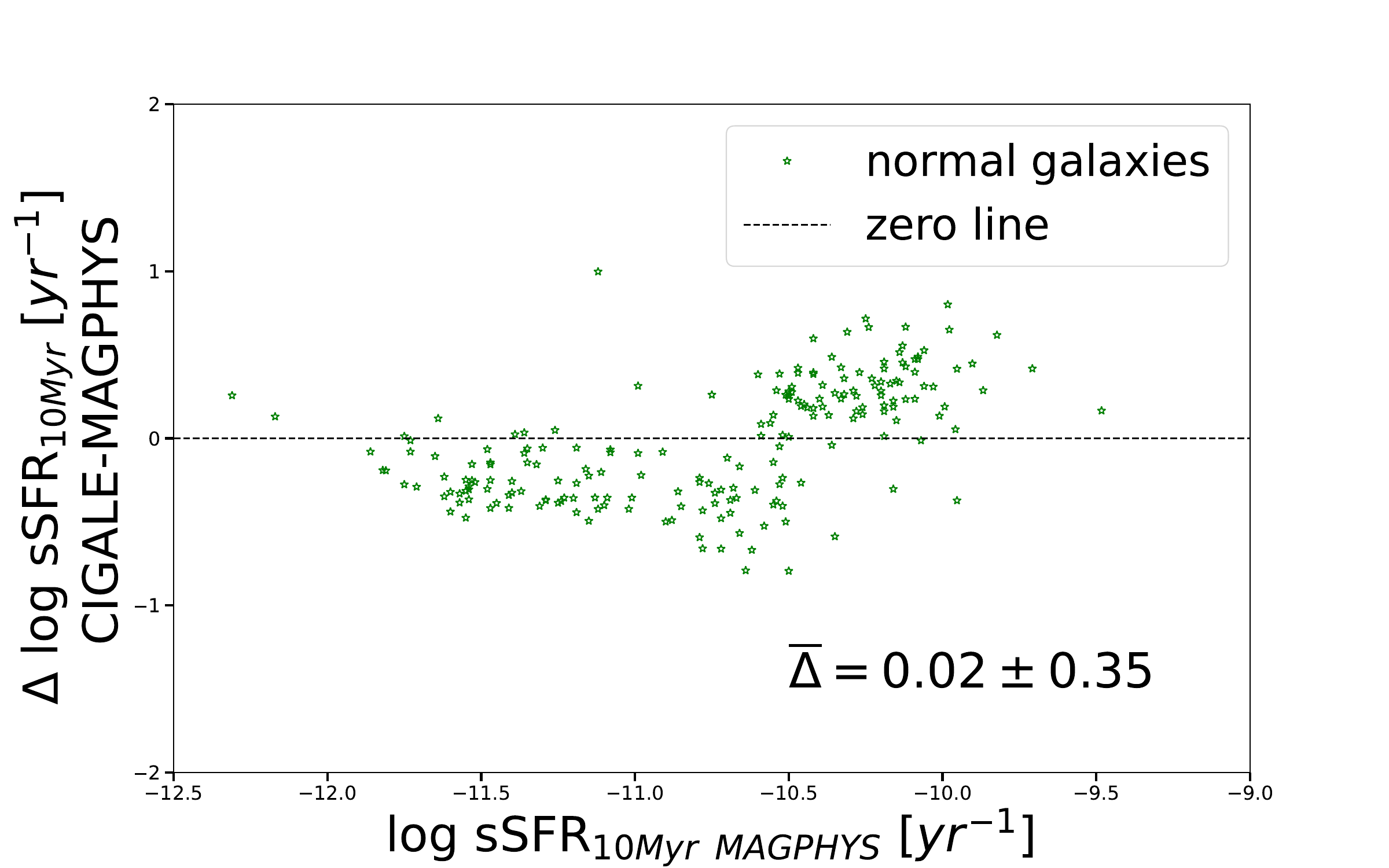}}
    \subfigure[]{\includegraphics[width=0.48\textwidth]{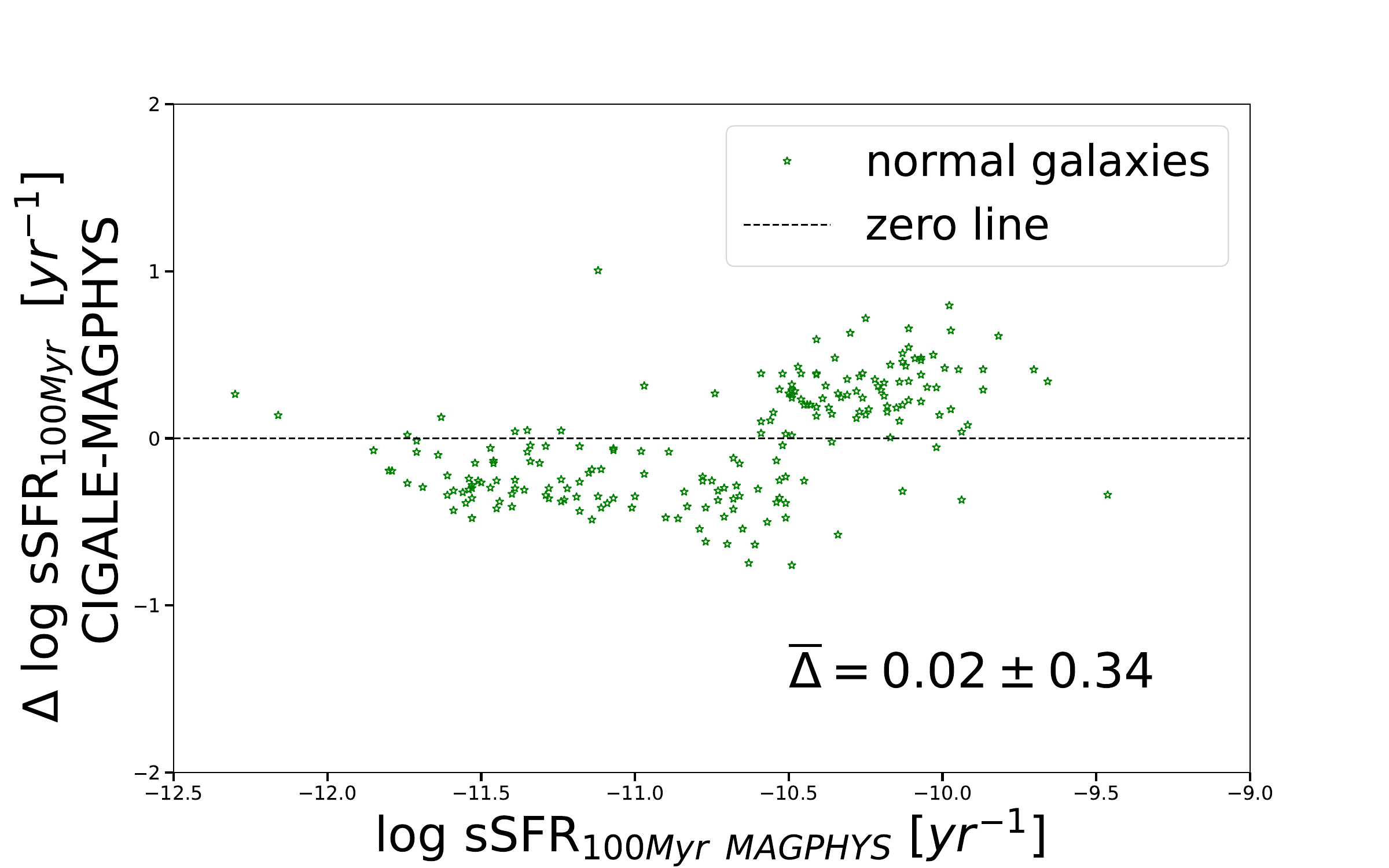}}
\caption{Differences ($\Delta$) between \textsc{cigale} and \textsc{magphys} estimates for normal galaxies  (without AGN component) for sSFR averaged over the past (a) 10 and (b) 100~Myr (test B).
The remark from Fig.~\ref{fig:comparisons_gals_withoutAGN} caption applies to these panels as well.}
\label{fig:comparisons_SFR_averages}
\end{figure*}

\textsc{magphys} and \textsc{cigale} also provide estimates for the SFR averaged over the past 10 and 100~Myr. 
\textsc{magphys} also provides estimates over the past 1000 and 2000~Myr.
Both codes measure the contributions from old stellar populations. 
In Fig.~\ref{fig:comparisons_SFR_averages_quasars} and~\ref{fig:comparisons_SFR_averages} we compare the results from \textsc{cigale} and \textsc{magphys} for the SFR averaged over the past 10 and 100~Myr. 
Similar to SFR estimates, we find that \textsc{magphys} overestimates SFR averages over the past 10~Myr but less so when averaged over the past 100~Myr.

In this work, for quasar host galaxies, we will use \textsc{cigale} derived values for SFR, sSFR, stellar masses and the SFR averaged over 10 and 100~Myr, as well as the contribution from older stellar populations. 
We will use the \textsc{magphys} values for the SFR averaged over the past 1000 and 2000~Myr as \textsc{cigale} does not provide these values. 
However, given the relative similarity of \textsc{cigale} and \textsc{magphys} results for the SFR averaged over the past 100~Myr and the older stellar populations, we regard this choice as permissible.

A thorough comparison of the results of both codes has been carried out by \cite{Hunt_2019} who also present a tabular version of similarities and differences between codes (e.g., assumptions in fits are given in their Table 1). 
Their results show that both codes yield similar estimates of parameters such as stellar masses, SFRs, and SFHs within 0.1 dex, i.e., within the typical photometric errors. 
This is confirmed in our comparison of \textsc{magphys} and \textsc{cigale} estimates of these parameters for a single set of normal galaxies , i.e., from test B. 
As in \cite{Hunt_2019}, we find a tendency for \textsc{cigale} to overestimate SFRs for very low star formation levels (also see more recent comparison for GAMAnear galaxies by \citealt{Paspaliaris_2023}).

We observe that in our dataset both methods also appear to yield similar results. 
The average differences $\Delta {\rm SFR_{CIGALE-MAGPHYS}}$, $\Delta {\rm sSFR_{CIGALE-MAGPHYS}}$, and $\Delta {\rm mass_{CIGALE-MAGPHYS}}$ are (0.03, 0.00, and 0.03). 
Almost all of the comparison normal galaxies lie between 0.2~dex of the \textsc{cigale} value in SFR, sSFR, and stellar mass (Fig.~\ref{fig:comparisons_gals_withoutAGN}). 
The similarity between SFR, sSFR, and stellar mass estimates based on \textsc{magphys}/\textsc{cigale} fits is further highlighted by the median differences of the property per 1 dex bin in each case (red dots in Fig.~\ref{fig:comparisons_gals_withoutAGN}).
There is a trend for \textsc{cigale} to overestimate SFR at low values, which is typical of quiescent galaxies (cf. \citealt{Hunt_2019}). 
The differences between values from \textsc{cigale} and \textsc{magphys} are significantly smaller than the histogram bins we use in subsequent figures. 
In the remainder of the analysis, we will use \textsc{magphys} values for normal galaxies in the CSNG, as these estimates are already provided by the GAMA survey for all 200 sets of comparison galaxies (while running \textsc{cigale} is found to be computationally expensive).

\subsubsection{Statistical correction for SFR values between \textsc{cigale} and \textsc{magphys}}
\label{sec:statcorrection}

In this work two SFR estimators are used - one based on \textsc{cigale} for the quasar sample because it includes AGN contribution and the second one from GAMA survey based on \textsc{magphys} for the control sample. 
We introduce a statistical correction to bring the \textsc{cigale} values to the \textsc{magphys} baseline.
Thus, any discrepancies between the two estimators are removed and consequently do not affect the results. 
Any systematic errors are also reduced.

To obtain the calibration function, we fit a Chebyshev polynomial to the SFR parameter offsets vs its \textsc{cigale} estimate, which is then used to obtain the corrected parameter value.

\begin{equation}
    log_{10}X_{corr} = log_{10}X_{CIGALE} - f_{Cheb}(log_{10}X_{CIGALE}) \, 
\end{equation}

where X stands for any SFR parameter, 
$f$ is the statistical correction function for the offsets, and 
X$_{corr}$ is the parameter value corrected to the \textsc{magphys} baseline.
This is applied to all SFR parameters of the quasar sample: SFR, sSFR, time-averaged SFR.
This correction is minor and doesn't substantially change the results.
The analysis uses the code described in the Software section.

\subsubsection{Main ingoing assumptions for \textsc{cigale} and \textsc{magphys}}
\label{sec:mainassumptions}

For completeness, we highlight the main assumptions and sources of uncertainties. 
Uncertainties in the SED estimates arise from the input photometry data.
There is no difference in input for the models between \textsc{magphys} and \textsc{cigale} in this work.
GAMA \textsc{magphys} SED fits are based on the LAMBDAR photometry. 
We used the same panchromatic data from LAMBDAR photometry for \textsc{cigale} fits.
The UV and IR survey photometry data in GAMA often has larger associated uncertainties compared to the measurements within the optical wavelengths. 
GAMA SED fits are based on a parameter space chosen to cover all the galaxies in the survey, not just the bright and low-\textit{z} galaxies.
As we are looking only at a subset of galaxies from the GAMA survey, our parameter space with SED fits with \textsc{CIGALE} performed here is tailored.

Both \textsc{magphys} and \textsc{cigale} adopt a computationally efficient energy balance approach to derive SEDs coupled with a Bayesian estimator to recover the full posterior probability distribution of galaxy physical parameters. 
The application of \textsc{magphys} to GAMA data is described in \cite{Driver_2018}, note that GAMA archival fits were performed on the full GAMA catalog, not focused on the low-$z$ sample of bright galaxies. 
There are uncertainties associated with models in each module of the SED program \citep[e.g.,][]{Conroy_2013}, which need to be remembered when interpreting results.
Both codes (as well as others) rely on a given SFH defined by an assumed IMF (\citealt{Chabrier_2003} for both codes) applied to a matrix of single age stellar populations (\citealt{Bruzual_2003} for both codes), see Table~\ref{tab:settings}. 
However, \textsc{cigale} adopts a single (solar) metallicity and an SFH based on a delayed parametrization approach as described by \cite{Ciesla_2015}, whereas \textsc{magphys} adopts an exponentially declining SFH, with SSPs having varying metallicities with random bursts to mimic realistic histories. 
However, there is no AGN module in \textsc{magphys}. 
The \textsc{magphys} research team is developing the inclusion of AGN but it is not yet finished.
We stress, however, that despite the underlying assumptions, the SED-based analysis generally produces good estimates of stellar population parameters for galaxies in general, as described in the relevant documentation papers for both codes.

\begin{deluxetable*}{llc}
\tablecaption{Comparison of the main ingoing assumptions between \textsc{cigale} fits and GAMA survey fits with \textsc{magphys} SED code. \label{tab:settings}}
\tablehead{
\colhead{Module} & \colhead{\textsc{magphys} fits in GAMA \citep{Driver_2018}} & \colhead{Contrast with \textsc{cigale} fits in this study}}
\startdata
IMF & \cite{Chabrier_2003} & same \\
SFH & Single (exponentially decaying) SFH  & different \\
SSP & BC03 models \citep{Bruzual_2003} & same \\
Dust Attenuation Law & \cite{Charlot_2000} & different \\
Metallicity & a wide range of values & a single value
\enddata
\tablecomments{
Column (1): Name of the component module.
Column (2): Assumed law and/or parameter values from the GAMA archival data, which is based on \textsc{magphys} SED tool.
Column (3): Comment on whether there are any differences when compared to the \textsc{cigale} SED fits performed in this work.
}
\end{deluxetable*}

Previous studies, such as \citealt{Stone_2021} and \citealt{Bettoni_2023}, were able to assess only a few parameters through optical spectroscopy, such as SFR, sSFR, and stellar mass estimates for quasar hosts based on emission line analysis. 
The SED analysis provides a larger set of parameter estimates which include star formation history parameters and AGN-related parameters. 
While future surveys are planned to get spectroscopy data for a large number of quasars, currently, GAMA survey already hold the photometry data for SED-based analysis.

\section{Results}
\label{sec:results}

In this section we present the distributions of star formation properties of low-$z$ quasar sample, obtained through the \textsc{cigale} SED tool (\S~\ref{sec:results_sfr}).
Moreover, various star formation history parameters are compared between the quasar sample and the CSNG (\S~\ref{sec:results_sfh}).

\subsection{Comparing star formation rate distribution of quasar sample with normal galaxies}
\label{sec:results_sfr}

\begin{figure*}[htb]
    \centering
    \includegraphics[width=\hsize]{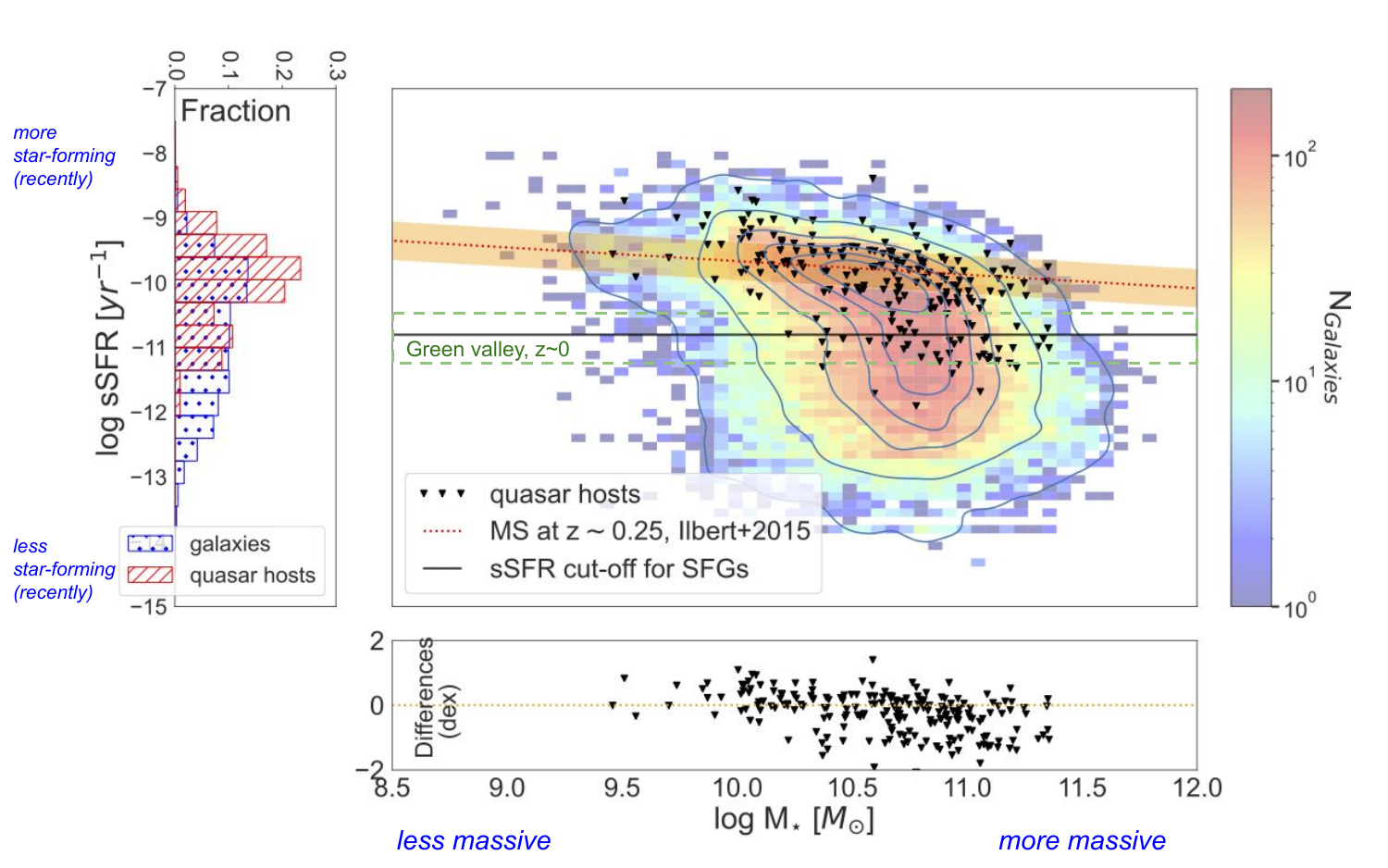}
    \caption{
    \textit{Top left} Distribution of sSFR for quasars and comparison galaxies.
    \textit{Top right} Stellar mass vs. sSFR for quasars (black triangles) and for comparison galaxies (contours and heatmap).
    The contour lines enclose 1/10/30/50/70/90 \% of comparison galaxies.
    The heatmap colors go blue to red corresponding to low to high values of comparison galaxy population counts, as indicated by the colorbar.
    The main sequence (MS) relation for SFGs from \cite{Ilbert_2015} is given here in orange dotted line, where the orange shaded area indicates the typical $\sigma$ value of 0.3~$dex$ \citep{Katsianis_2019}.
    The black solid line indicates the cutoff between SFGs and passive galaxies, $log(sSFR)=-10.8\ yr^{-1}$.
    The redshift value for the MS was adopted to be the median redshift of the SFGs in this sample of comparison galaxies, as discussed in \cite{Stone_2023}.
    For reference, the green valley transition region is marked per \cite{Salim_2014} with a green dashed box (for \textit{z}$\sim$0).
    \textit{Bottom} Residuals which show the logarithmic difference (dex) between the quasar sSFR values and the sSFR values from the reference MS relation (orange dotted line), $\Delta (dex)= log_{10}(sSFR_{quasar}/sSFR_{ref})$.
    The plotted sSFR and stellar mass data for comparison galaxies come from the GAMA survey \textsc{magphys} SED analysis, while for quasar hosts from \textsc{cigale} SED fits. 
    For quasar hosts, the sSFR values from \textsc{cigale} were corrected to MAGHPYS baseline.
    }
    \label{fig:ssfr_mass_plot}
\end{figure*}

To demonstrate how the SFR of quasar hosts compares to that of normal galaxies, we map in Fig.~\ref{fig:ssfr_mass_plot} (right) the stellar mass vs. sSFR for the sample of quasar host galaxies and for a Monte Carlo realization of 200 samples of 205 normal galaxies. 
The majority of quasars, 80\%, lie above the cutoff between star-forming galaxies (SFGs) and passive galaxies, $\log(sSFR)=-10.8\ M_{\odot}$ yr$^{-1}$. 
(Note that $\sim$15\% of the quasar sample is radio-loud using limit of 1~mJy for 20~cm emission.)

\begin{figure*}[htb]
    \centering
    \subfigure[]{\includegraphics[trim=0 0.3cm 0 2cm, clip=true, width=0.47\textwidth]{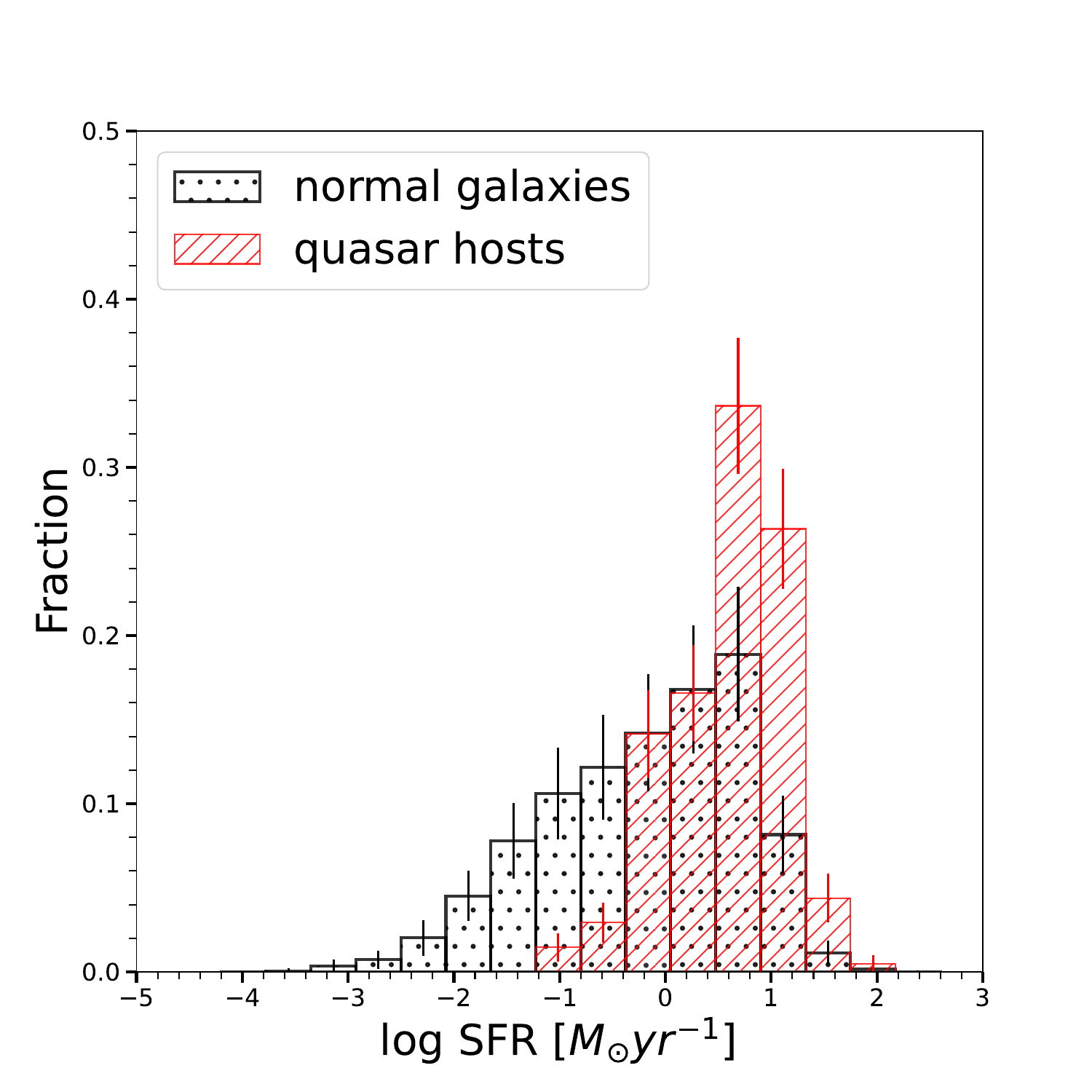}}
    \subfigure[]{\includegraphics[trim=0 0.3cm 0 2cm, clip=true, width=0.47\textwidth]{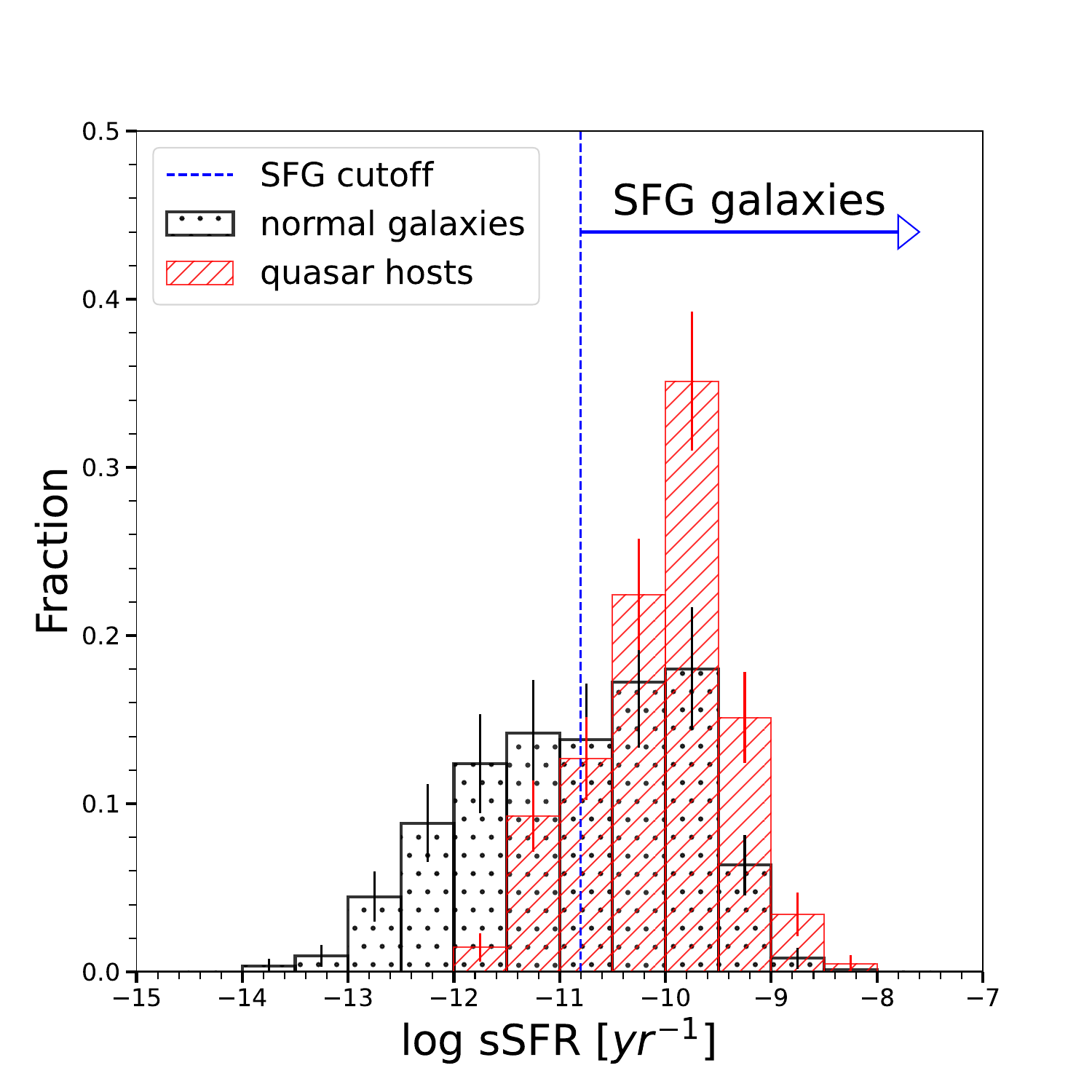}}
\caption{Histograms of SF properties with Poisson error bars: (a) SFR, (b) sSFR. 
Data for comparison galaxies are from GAMA archive, for quasar hosts from \textsc{cigale} SED fits. 
For quasar hosts, \textsc{cigale} SFR and sSFR estimates are corrected to \textsc{magphys} baseline and the error bar represents the Poisson error. 
For comparison galaxies, the plotted fraction in each bin is the average fraction from all 200 realizations, and the error bar represents the standard deviation.
In each case, the distribution of values for quasar sample lies in the peak of the normal galaxy distribution. 
However, there are few quenched AGN hosts.
}
\label{fig:histograms_sfr}
\end{figure*}

The comparison galaxies are chosen to lie in the same volume, have the same stellar mass and redshift as the sample of quasar host galaxies. 
We further contrast the two samples in Fig.~\ref{fig:histograms_sfr}, where we plot the histograms of SFR (a) and sSFR (b) for quasar host galaxies and the CSNG.

\subsection{Star formation history parameters of quasar hosts}
\label{sec:results_sfh}

\begin{figure*}[h]
    \centering
    \includegraphics[angle=-90, width=0.65\textwidth]{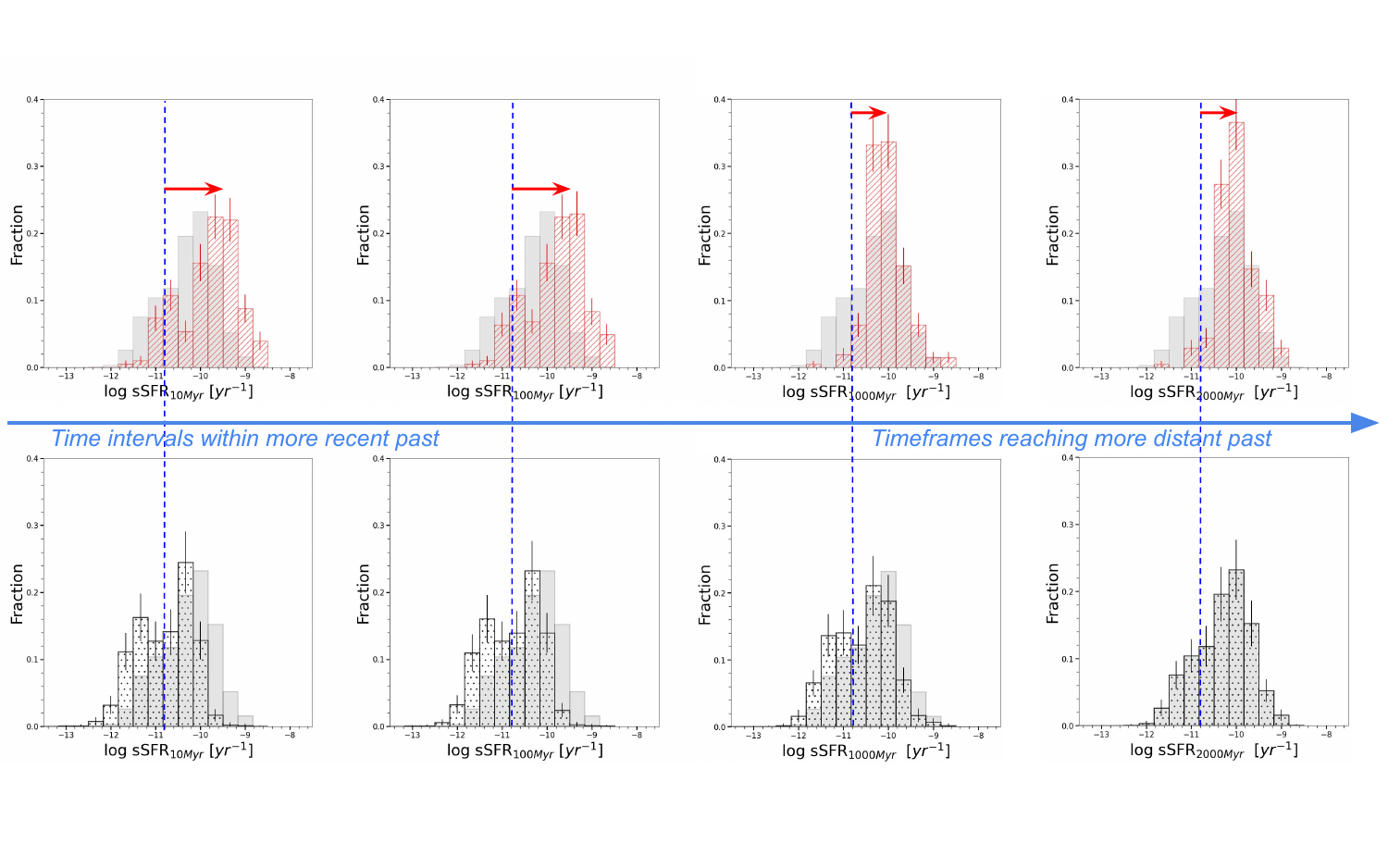}
    \caption{Histograms of sSFR averaged over 10, 100, 1000, and 2000~Myr for quasars (red hashed bars) and for normal galaxies (black dotted bars).
    For quasars, average sSFR estimates from \textsc{cigale} for 10~Myr and 100~Myr are corrected to \textsc{magphys} baseline.
    For visual contrast, the 2~Gyr distribution for normal galaxies is shown in shaded bars in each plot. 
    Error bars as in Fig.~\ref{fig:histograms_sfr}.
    The plots are arranged along the increasing timescale over which SFR is averaged.
    To highlight the comparison and the evolution of the distributions, the blue dashed line marks the sSFR=-10.8 yr$^{-1}$ line in all plots, while the red arrow on quasar plots shows the separation to the approximate peak of the distribution. 
    Thus, quasar hosts sample exhibits elevated SFR values at 100 and 10 Myr marks.
    Note that the SFR peak for quasar hosts is moved towards higher values in more recent timeframes.
    On the other hand, there is no evidence that the SFR of normal galaxies is changing with time. 
    Indeed it seems that the star-forming cloud is always in the same place even at high redshift.
    }
\label{fig:histograms_sfr_averaged}    
\end{figure*}

Star formation history indicators, such as SFR averaged over 10, 100, 1000~Myr can be employed to check how star formation evolves and compare the results with time-interval from quasar evolution models.
\textsc{cigale} provides estimates of the SFR averaged over the past 10 and 100~Myr. 
\textsc{magphys} also provides estimates over the past 1000 and 2000~Myr but, unfortunately (as we can see in Fig.~\ref{fig:comparisons_SFR_averages_quasars}) this neglects the influence of the quasar. 
In that figure, we see that the effect of the AGN is less important when averaged over longer epochs, so we also present \textsc{magphys} data (for SFR averaged over 1000 and 2000~Myr) with the caveat that these estimates may be affected by AGN light (i.e., overestimated). 
This is shown in Figure~\ref{fig:histograms_sfr_averaged}.

\begin{figure*}[hbt]
\centering
\subfigure[]{\includegraphics[trim=0 0 0 2cm, clip=true, width=0.47\textwidth]{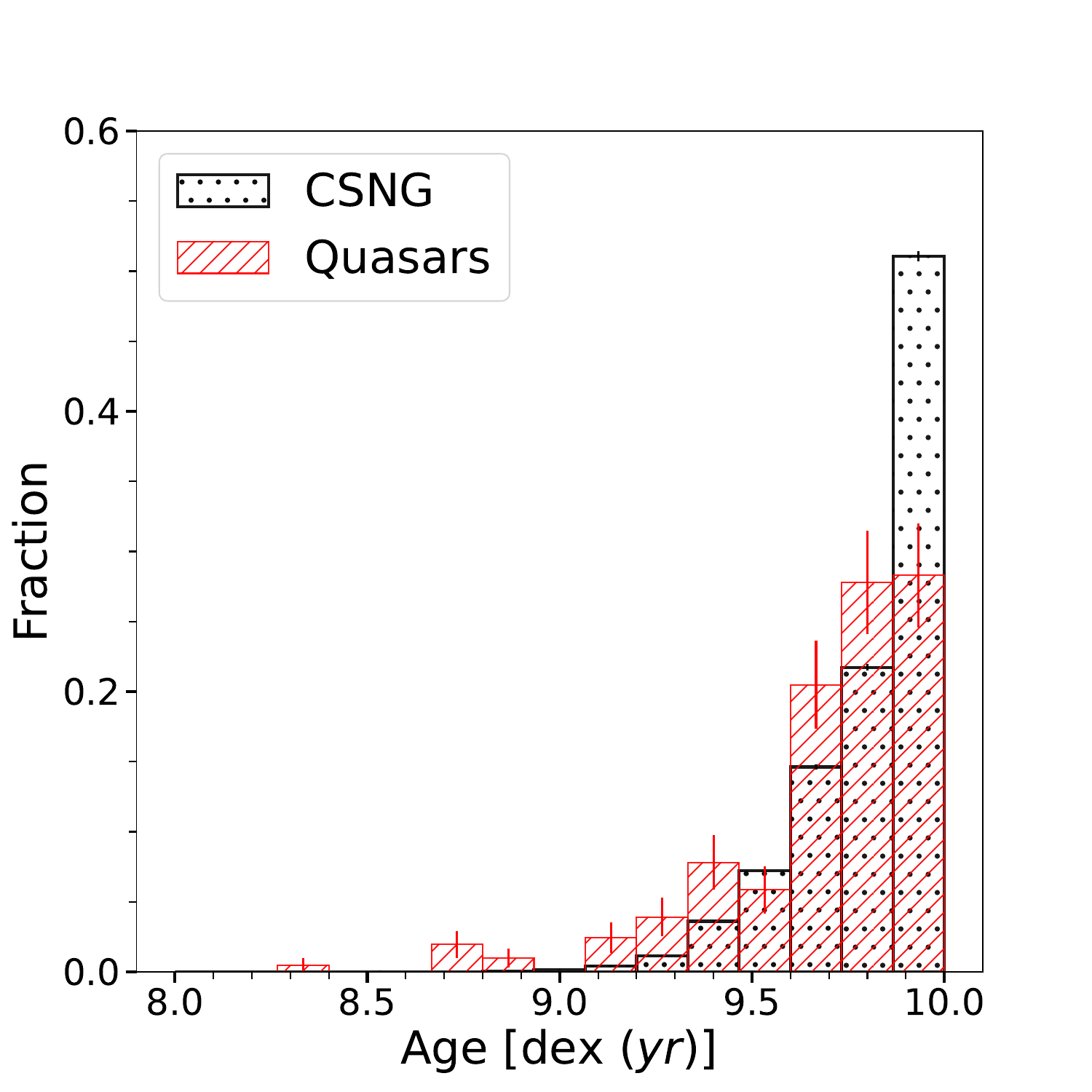}}
\subfigure[]{\includegraphics[trim=0 0 0 2cm, clip=true, width=0.47\textwidth]{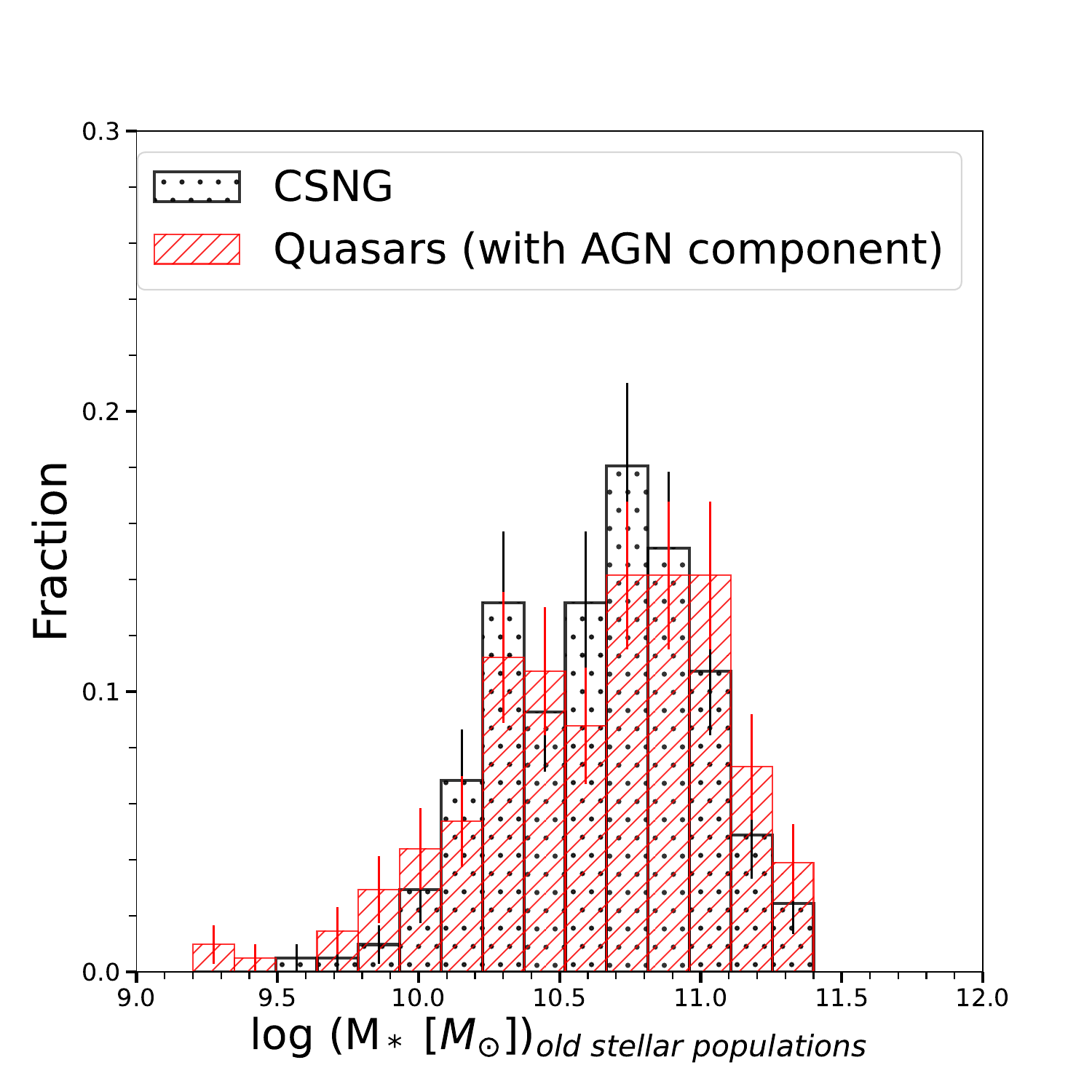}}\\
\subfigure[]{\includegraphics[trim=0 0 0 2cm, clip=true, width=0.47\textwidth]{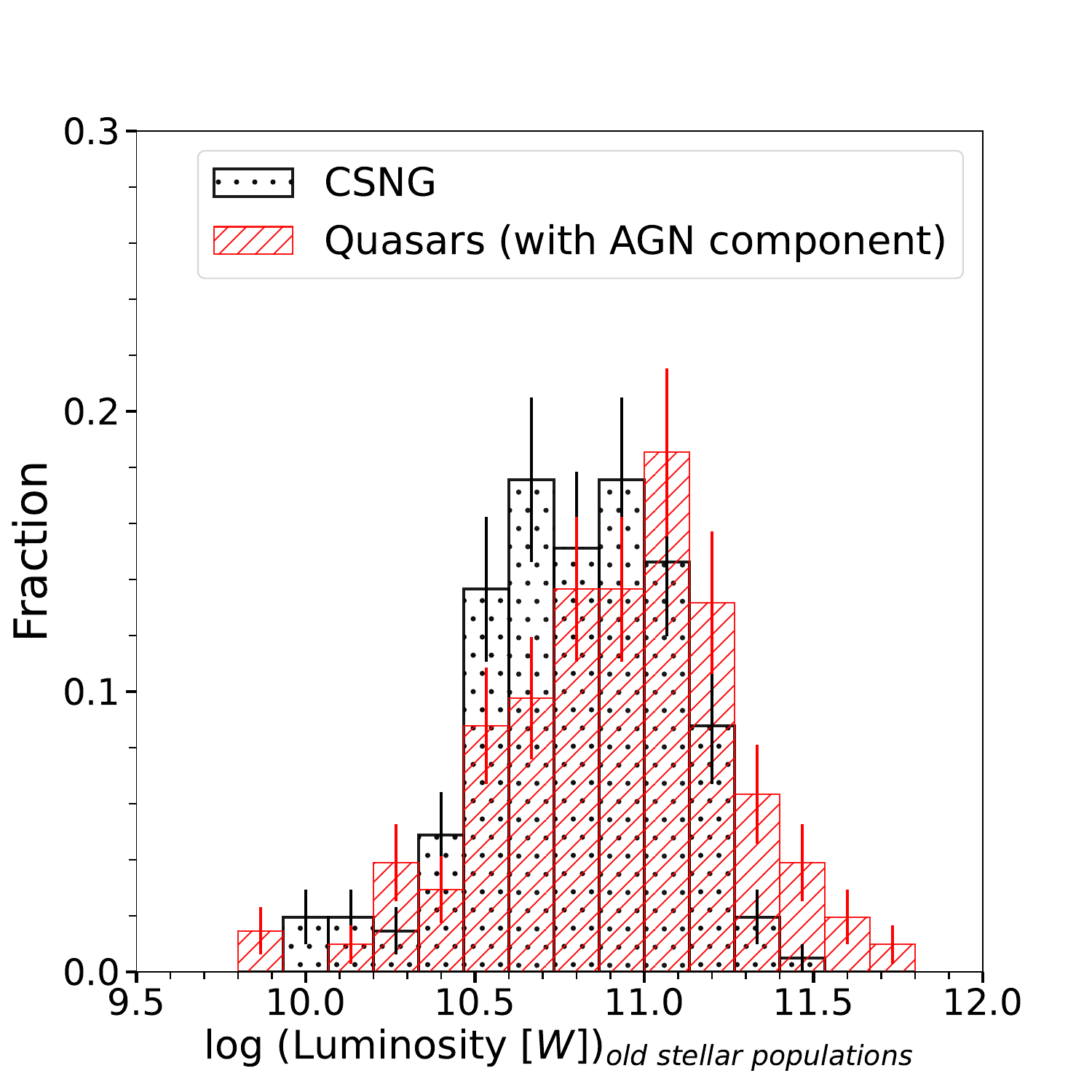}}
\caption{Histograms for old stellar populations for both the quasar sample (red hatched bars) and for the CSNG (black dotted bars). 
(a) Age of the oldest stars in the galaxy from \textsc{magphys} fits. 
(b) Stellar mass and (c) luminosity of old populations from \textsc{cigale} fits.
The error bars are as in Figure~\ref{fig:histograms_sfr}. }
\label{fig:histograms_old}
\end{figure*}

Further information about the star formation history is gleaned from the perspective of the old stellar population parameters.
We present in Fig.~\ref{fig:histograms_old} the distributions of age of the oldest stars, as well as the stellar mass and luminosity of old stellar populations.
The plots show no significant difference between CSNG and the quasar sample.

\section{Discussion}
\label{sec:discussion}

We interpret the results disclosed in \S~\ref{sec:results} here.

\subsection{Quasar host galaxies are mostly star-forming}

The distribution of quasar host galaxies in the stellar mass vs. SFR (sSFR) plot (Fig.~\ref{fig:ssfr_mass_plot}) is broadly consistent with that of normal SFGs, with a smaller fraction lying in the green valley or just below in the quenched region. 
This can also be observed in the histograms shown in Fig.~\ref{fig:histograms_sfr}. 
There are few to no quasars in the fully quenched region. 
However, the more quiescent objects can generally be seen to be more massive systems as in \cite{Smirnova_2022}.

This confirms that quasar host galaxies are generally similar to normal SFGs and are less represented among the quiescent population. 
Of course, this might not be the case for very powerful quasars and radio galaxies that are rarer and will only rarely be present in volumes the size of the GAMA survey.

When we compare with the distribution for SFGs we note that there is a smaller fraction of quasar host galaxies in the green valley, with a few quasars lying at the top end of the quiescent galaxies distribution. 
This suggests that star formation may be quenched in otherwise normal SFGs.
An alternate hypothesis to explain the deficit of quasar host galaxies in the green valley would be that quasar host galaxies initially lie in the green valley and undergo increased SFR later. 
In order to test this we can look at the averaged SFR over the past 10 and 100~Myr and then compare the star formation timescales to the expected quasar lifetimes, discussed next.

\subsection{Quasar hosts show increase in star formation at recent timescales}
It is interesting to consider the timescale for star formation and compare with the expected quasar lifetimes. 
In the model of \cite{Hopkins_2006} quasars are associated with increased star formation over timescale of $\sim$10~Myr followed by quenching once the quasar becomes optically bright (see also \citealt{Morey_2021}). 

The SFR and sSFR distributions for non-AGN galaxies (CSNG) appear to be relatively similar at all times, suggesting that these objects have undergone a comparatively quiet evolution since $z \sim 0.3$, as demonstrated by the sequence of bottom panels in Figure~\ref{fig:histograms_sfr_averaged}. 
Similarly, this is the case for the SFR and sSFR distributions for quasar host galaxies for timescales of 1000 and 2000~Myr (top rightmost panels in Figure~\ref{fig:histograms_sfr_averaged}).

In contrast, there is evidence of an increase in SFRs in quasar host galaxies within the last 100~Myr and 10~Myr (top leftmost panels in Figure~\ref{fig:histograms_sfr_averaged}). 
We see that the SFR and sSFR distributions for quasar host galaxies are shifted towards higher values over more recent timescales, implying an association between an increase in star formation activity and nuclear activity.
Note that if the SFR averaged over 1000 and 2000~Myr is overestimated for quasar hosts by \textsc{magphys}, then the resultant increase in star formation is even more profound, and thus does not revert the conclusions of this work.

We stress that all quasar host galaxies, however, were forming stars on the star-forming MS within the last 2~Gyrs. 
In other words, quasar activity is associated with an increase in SFRs in already normal SFGs. 
Increased SFR is therefore associated with quasar activity, but the starburst entity is modest, similar to the episodic SFR encountered in normal spirals.

\subsection{Deficit of quasar host galaxies in Green Valley region}

Let's consider again the sSFR vs. stellar mass diagram (Fig.~\ref{fig:ssfr_mass_plot}), which shows that there is a relative deficit of quasar host galaxies (compared to SFGs and all galaxies) in the traditional green valley region and the quiescent region.
The distribution of SFRs for quasar host galaxies appears bimodal, with a small fraction of objects in the quiescent portion of the sSFR-stellar mass plot, but lying close to the sSFR limit for galaxies to be identified as star-forming. 
This hints that quasar host galaxies may have been more quiescent prior to the activation of the AGN and closer to the green valley, as argued by \cite{Povic_2012}. 
However, the older stellar populations of quasar host galaxies are not significantly different from those of normal galaxies (Fig.~\ref{fig:histograms_old}). 
Nevertheless, quasar host galaxies have always been star-forming over these timescales, lending support to models where nuclear activity is triggered by secular evolution.

\subsection{Implications to models linking SFR and AGN activity}

Our finding that most quasar host galaxies lie on or close to the star-forming MS is consistent with the observations by \cite{Smirnova_2022}, where no significant effect was found on the SFRs of quasar host galaxies. 
Coupled with the lack of quiescent galaxies among quasar hosts (although there may be radio galaxies that do not fulfill the criteria for inclusion by \citealt{Gattano_2018}), this is consistent with recent studies suggesting that there is no significant effect from feedback by active nuclei \citep{Elbaz_2011,Bongiorno_2012,Harrison_2012,Balmaverde_2016,
Leung_2017,Woo_2017,Shangguan_2018,Scholtz_2020}. \cite{Zhuang_2022} also find vast majority of quasar host galaxies are star-forming, while \cite{Koutoulidis_2022} claim that 3/4 of their X-ray selected AGN lie on or above the star-forming MS, but \cite{Shangguan_2020,Shangguan_2020b} and \cite{Xie_2021} argue that the SFRs of quasar host galaxies in their Palomar-green sample approximate those of starbursts. \cite{Mountrichas_2024a,Mountrichas_2024b} also find that only quasars with high X-ray luminosity have higher SFRs than comparable non-active galaxies.

About 1/4 of quasar hosts are transitioning from star-forming to the quenched population of galaxies; they all have [OII] emission but this is not securely associated with current star formation as the spectra may be contaminated with emission from the nuclear region. 
This is an indication that nuclear activity may continue for a considerable time after star formation is quenched (as in \citealt{Rembold_2017,Sanchez_2018}). 

\cite{Goto_2006} claims that about 5\% of quasars show post-starburst signatures.
Post-starburst signatures are more recently observed also in the off-center optical spectra of low-\textit{z} Type~I quasar hosts \citep{Bettoni_2017,Stone_2021}.
In this case, nuclear activity continues for a considerable time after quenching, since the presence of post-starburst features implies that star formation ceased a few hundred Myr ago. 
This is in contrast to the simple model where quasars drive rapid quenching by feedback and then dwindle.

However, we can also show that most of the increase in SFRs has taken place over the past $\sim$100~Myr, while all quasar host galaxies are seen to be normal SFGs over the past 1--2~Gyr. 
None of the quasar hosts was originally quiescent. Therefore this argues that the AGN activity is related to an increase in SFRs in otherwise normal galaxies. 
Deep integral field unit and long slit spectra of nearby AGNs show that peak AGN activity occurs $\sim$50–-200~Myrs after the onset of star formation (\citealt{Davies_2007,Riffel_2021} see also \citealt{Schawinski_2009,Wild_2010}). 
Similar timescales are seen in \cite{Rembold_2017} and \cite{Sanchez_2018}. 
This is consistent with our findings that star formation in quasar host galaxies has increased over the past 100~Myr. 

The star-forming region may provide the fuel for the SMBH and allow such fuel to reach the SMBH via increased turbulence. 
Analytical models, numerical simulations, and hydrodynamical simulations predict that stellar winds and supernovae will enhance the SMBH mass accretion by injecting turbulence into the gas disk  (e.g., \citealt{Wada_2002,Schartmann_2009,Hobbs_2011}) while several authors have now reported a correlation between the black hole accretion rate and increasing SFRs \citep{Mullaney_2012,Chen_2013,Harris_2016,Lanzuisi_2017,Zhuang_2020,Zhuang_2021}. 
This association between quasars and star formation need not be causal: it may simply reflect the availability of a large gas supply, some of which is driven to the nucleus to fuel the AGN \citep{Jarvis_2020,Shangguan_2020,Yesuf_2020}.

Based on what we see in Fig.~\ref{fig:histograms_sfr}--\ref{fig:histograms_old} we suggest that the recent increase in star formation and AGN activity may have occurred among galaxies in the green valley, where quasar host galaxies seem to be less frequent.
These galaxies may experience an increase in star formation and this may then induce quasar activity, as argued by \cite{Povic_2012,Charlton_2019,Lin_2022} where X-ray selected AGN are observed preferentially in the green valley region (i.e., at lower SFRs). The SFR increase may be episodic as is observed in many spiral galaxies rather than due to mergers and interactions, although these could certainly play a role as in our Galaxy \citep{Ruiz-Lara_2020}. 
Gas from young stars may then feed AGN activity, explaining the observations in Fig.~\ref{fig:histograms_sfr}--\ref{fig:histograms_old}. This may also explain the different claims about the correlation between star formation, quenching and quasar activity, if samples are chosen in different portions of this life cycle. 

It is unclear what mechanisms may produce an increase in the SFR and fuel the AGN. Major mergers are clearly able to induce strong tidal torques and drive gas and dust to the center, but they are likely to be important only for the most powerful AGN. There is no strong evidence that quasar host galaxies are more morphologically disturbed than normal galaxies of the same luminosity \citep{Grogin_2005,McKernan_2010,Cisternas_2011,Boehm_2012,Kocevski_2012,Villforth_2014,
Villforth_2019,Marian_2019,Shah_2020}. It has been shown that quasar host galaxies favor relatively low density environments \citep{Wethers_2022} and do not have significantly more or fewer companions \citep{Stone_2023}, while the star formation properties and stellar populations of galaxies neighboring quasar hosts are consistent with those of normal galaxies of the same mass \citep{Bettoni_2017,Bettoni_2023,Stone_2021,Stone_2023}. 

This leaves minor mergers and secular evolution, such as bars and spiral arms \citep{Shlosman_1989,Hopkins_2010} as possible triggers (this conclusion stands even though GAMA has few massive cluster environments). 
These suffice to fuel most of AGN activity (e.g., \citealt{Ho_2009,Cisternas_2011,Villforth_2017,Zhao_2022}).
The timescales we find here are long enough to favor such gentler and slower processes. 

\section{Conclusions}
\label{sec:conclusions}

In this work we investigated the SFH of low-redshift (0.1$<$z$<$0.35) Type I AGN within the GAMA volume and compared these objects to non-active galaxies in the same survey matched in redshift and stellar mass. 
Our main conclusions are as follows.
\begin{enumerate}
    \item Most (80\%) quasars lie on the star-forming MS for normal galaxies and 20\% are hosted in quenching or quenched systems. 
    \item The SFHs of these quasar host galaxies show that they have generally been SFGs over the past 2~Gyr but have experienced a modest star formation increase (by a factor of 2--3) over the past 100~Myr. 
    \item These observations support models where star formation feeds quasar activity, and its cessation starves the SMBH of gas and dust, leading to its undergoing a quiescent state. 
    \item We conclude that the correlation between the SMBH mass and the bulge mass may not have a causal relation and that AGN activity, at least in the local universe, is mainly triggered by secular processes within normal galaxies.
\end{enumerate}
Extending this analysis to a much larger quasar sample is necessary to more definitively validate the conclusions of this work.
For example, future work is planned with the quasars surveyed by the Dark Energy Spectroscopic Instrument \citep[DESI,][]{DESI_2016}.

\section{Acknowledgements}
MBS acknowledges the Finnish Cultural Foundation grant numbers 00220968 and 00231098, the Academy of Finland grant 311438, and FINCA funding.
MBS acknowledges useful email discussions with Pekka Heinämäki, Johanna Hartke, Véronique Buat, Denis Burgarella, Kostas Kouroumpatzakis, Elina Lindfors, Talvikki Hovatta, and with GAZPAR support personnel for \textsc{CIGALE}, operated by CeSAM-LAM and IAP, as well as support with GAMA survey questions from Ivan Baldry, Simon Driver, Jonathan Loveday, and Alister Graham, as well as with James Stone and Kaj Wiik for help with CIGALE installation; plus science discussions with FINCA community and feedback from referee.

GAMA is a joint European-Australasian project based around a spectroscopic campaign using the Anglo-Australian Telescope. The GAMA input catalogue is based on data taken from the Sloan Digital Sky Survey and the UKIRT Infrared Deep Sky Survey. Complementary imaging of the GAMA regions is being obtained by a number of independent survey programmes including GALEX MIS, VST KiDS, VISTA VIKING, WISE, Herschel-ATLAS, GMRT and ASKAP providing UV to radio coverage. GAMA is funded by the STFC (UK), the ARC (Australia), the AAO, and the participating institutions. The GAMA website is http://www.gama-survey.org/ .

This research has made use of the Spanish Virtual Observatory (https://svo.cab.inta-csic.es) project funded by MCIN/AEI/10.13039/501100011033/ through grant PID2020-112949GB-I00; the NASA/IPAC Extragalactic Database (NED), which is funded by the National Aeronautics and Space Administration and operated by the California Institute of Technology; StackOverflow platform, the VizieR database \citep{Ochsenbein_2000}; Cosmology Calculator \citep{Wright_2006}, Seafile file collaboration platform.

\vspace{5mm}
\facility{Anglo-Australian Telescope (AAT)}

\software{astropy \citep{Astropy_2013,Astropy_2018},
            matplotlib \citep{Hunter_2007},
            numpy \citep{Harris_2020}, 
            python \citep{Python_2009},
            seaborn \citep{Waskom_2021},
            scipy \citep{Virtanen_2020}, 
            code from \cite{Stone_2017}.  
            Open access is released for the code associated to this part, as minimal code example \citep{Stone_2025_repA} and as the detailed calculations \citep{Stone_2025_repB}.
          }
% other software packages?

\bibliography{references}{}
\bibliographystyle{aasjournal}

\listofchanges

\end{document}